\documentclass[aps,prb,twocolumn,superscriptaddress,showpacs,floatfix]{revtex4-2}
\usepackage{amsmath,amssymb,amsfonts,float,graphics,epsfig,epstopdf,color,verbatim,tabularx,bm,multirow,appendix,hyperref}
\usepackage{amsmath}
\usepackage{amssymb}
\usepackage{mathtools}
\usepackage{graphicx}
\usepackage{bm}
\usepackage{hyperref}
\usepackage{url}
\usepackage[utf8]{inputenc}
\usepackage{subfigure}
\usepackage{slashed,bbm}
\usepackage{graphics,psfrag,epsfig}
\usepackage{dsfont}
\usepackage{setspace}
\usepackage{wasysym}
\usepackage{slashed}

\usepackage{hyperref}
\usepackage{xcolor}
\definecolor{dark-red}{rgb}{0.4,0.15,0.15}
\definecolor{dark-blue}{rgb}{0.15,0.15,0.4}
\definecolor{medium-blue}{rgb}{0,0,0.5}
\hypersetup{
colorlinks, linkcolor={dark-blue},
citecolor={dark-blue}, urlcolor={medium-blue}
}

\newcommand{\be}{\begin{equation}}
\newcommand{\ee}{\end{equation}}
\newcommand{\bea}{\begin{eqnarray}}
\newcommand{\eea}{\end{eqnarray}}

\newcommand{\hc}{\text{h.c.}}

\renewcommand{\i}{\text{i}}
\DeclareMathOperator{\tr}{tr}

\usepackage[normalem]{ulem}

\begin{document}

\title{Fermionic skyrmions and bosonization for a Gross-Neveu transition}

\author{Xiao Yan Xu}
\affiliation{Key Laboratory of Artificial Structures and Quantum Control (Ministry of Education), School of Physics and Astronomy, Shanghai Jiao Tong University, Shanghai 200240, China}
\author{Tarun Grover}
\affiliation{Department of Physics, University of California at San Diego, La Jolla, California 92093, USA}

\begin{abstract}
We investigate a 2+1-D interacting Dirac semimetal with onsite flavor SU(2) symmetry. Topological considerations imply that the skyrmions in the flavor-symmetry-breaking phase carry electron quantum numbers,  motivating a dual bosonized low energy description in terms of  two complex scalars coupled to an abelian Chern-Simons field. We propose that the transition between a nearby Chern insulator and the flavor symmetry-broken phase is a bicritical point in the bosonized description, and also suggest that the Gross-Neveu-Heisenberg (GNH) transition between the Dirac semimetal and the flavor symmetry-broken phase is a tricritical point. Heuristically, the dual description corresponds to the gap closing of fermionic skyrmions. We discuss implications and potential issues with our proposal, and motivated from it, perform extensive unbiased Determinantal Quantum Monte Carlo (DQMC) simulations on a lattice regularized Hamiltonian for the GNH transition, extending previously available results. We compare DQMC results with the estimates in the proposed dual to available perturbative renormalization group results. We also numerically demonstrate the presence of fermionic skyrmions in the symmetry-broken phase of our lattice model. 

\end{abstract}

\maketitle

\section{Introduction} \label{sec:intro}
Quantum numbers associated with solitonic textures and topological defects are crucial in a wide range of phenomena, including fractionalization, quantum criticality, and the determination of exchange statistics for emergent excitations \cite{finkelstein1968connection, jackiw1976solitons,su1979solitons,goldstone1981fractional, Wilczek1983, skyrme1961non,witten1983current, lee1990boson, sondhi1993skyrmions, abanov2000theta, senthil2004deconfined, senthil2004quantum}. These quantum numbers also play a role in dualities relating seemingly different theories that have led to new connections between exotic quantum criticality, interacting topological insulators, and compressible quantum Hall systems \cite{peskin1978mandelstam,dasgupta1981phase, chen1993mott, senthil2006competing, maldacena2013constraining, giombi2012chernsimons, aharony2012correlation, aharony2016baryons, seiberg2016dualityweb, hsin2016level, karch2016particlevortex, karch2016more, barkeshli2014continuous, son2015composite, chong2016half, chong2015dual, metlitski2016particlevortex, metlitski2015s, xu2015selfdual, mross2016explicit, raghu2018exact,  wang2017deconfined, qin2017duality,senthil2019duality}. In this paper, we explore a setup where skyrmions of an SU(2) flavor symmetry breaking phase carry electron quantum numbers. Inspired by this observation, we consider a proposal for a dual of a Gross-Neveu-type \cite{Gross74} phase transition between the ordered phase and a Dirac semimetal. This phase transition can be realized in a lattice-regularized Hamiltonian that can be simulated without the fermion sign problem \cite{Lang2019}. We perform detailed Quantum Monte Carlo (QMC) simulations on the corresponding Hamiltonian, obtaining new results for the universal scaling exponents relevant to our duality proposal.  A notable feature of the field theory we investigate is it cannot arise in a purely local two-dimensional lattice model with time-reversal symmetry and on-site flavor symmetry. This characteristic is evident in the sign-problem-free lattice model that we simulate.

The starting point of our discussion is a Dirac semimetal in 2+1-D with two flavors of a two-component Dirac spinor. Interactions can lead to the spontaneous breaking of the SU(2) flavor symmetry down to U(1). Using standard arguments, this transition can be described by the so-called `Chiral Gross-Neveu-Heisenberg' (GNH) field theory, where electrons are coupled to a fluctuating O(3) order-parameter \cite{Gross74,rosenstein1993critical,herbut2006interactions,herbut2009relativistic,sorella2012absence,assaad2013pinning,janssen2014antiferromagnetic}. As already mentioned, a noteworthy aspect of the symmetry-broken phase in our model is that the skyrmions of the order parameter carry the same quantum numbers as the microscopic electrons \cite{abanov2000theta, Abanov2000}. This suggests a physical picture where, as one approaches the transition from the ordered side, the gap to skyrmions closes at the transition, resulting in the semimetal phase. The low-energy theory in the ordered phase can be reformulated as a Chern-Simons-matter theory  where a two-component complex scalar is coupled to a dynamic U(1) Chern-Simons gauge field whose flux corresponds to the skyrmion density \cite{Wilczek1983}. This motivates us to explore the phase diagram of our model by tuning the parameters in a Chern-Simons-matter theory whose field content is similar to the aforementioned field theory deep within the ordered phase. We find that the mass change of the complex scalar describes a transition between the flavor-symmetry broken phase and a Chern insulator. This motivates us to suggest that the GNH critical point where the three phases, the Dirac-semimetal, the Chern insulator, and the ordered phase meet, is dual to an interaction-tuned tricritical point in this Chern-Simons-matter theory. We discuss implications and potential issues with such a proposal, and motivated from it, compare our QMC results with available results from large-$N$ expansions on the tricritical theory.

From a numerical standpoint, the lattice-regularized GNH model we employ was originally introduced and studied by L\"auchli and Lang in Ref.\cite{Lang2019}. However, connections to any potential duality or topological aspects, such as the quantum numbers of skyrmions, were not considered. Ref.\cite{Lang2019} obtained scaling exponents of several operators corresponding to the GNH transition. Inspired by the proposed duality, we will provide universal exponents of several additional operators, such as the two-point correlation function of skyrmion density, fermion mass, and electron pairs.

Since the exchange statistics of skyrmions plays a key role in the proposed duality, we will also implement a lattice regularized numerical demonstration of the skyrmions' fermionic statistics. The main idea is to measure the Berry's phase associated with a process that generates a skyrmion-antiskyrmion pair from the vacuum, rotates one of them by $2 \pi$, and then annihilates the pair back into the vacuum \cite{Wilczek1983}.

The paper is organized as follows: In Sec.\ref{sec:latticemodel} we introduce the lattice regularized model which exhibits the GNH transition. In Sec.\ref{sec:gnhduality} we consider a bosonized description of various phases and phase transitions in terms of a Chern-Simons-matter theory. In Sec.\ref{sec:qmc_results} we discuss our QMC results in light of the dual formulation. In Sec.\ref{sec:rotation} we provide a numerical demonstration of the fermionic exchange statistics of the skyrmions. We conclude in Sec.\ref{sec:discuss} with a discussion of our main results, potential issues and future directions.

\section{Phase diagram of a lattice regularized  GNH model} \label{sec:latticemodel}

The theory we are interested in involves two flavors of two-component Dirac fermions in 2+1-D. One way to realize such a theory is by considering spinless fermions at half-filling with opposite sublattice hopping on a honeycomb lattice. In the absence of interactions, one obtains two Dirac nodes in the momentum space, whose low energy degrees of freedom correspond to the two flavors. However, if one desires a theory where the flavor symmetry acts locally in real-space, then the most physical way to realize the theory of our interest is at the 2+1-D boundary of a 3+1-D C-I class topological superconductor \cite{wang2014interacting}. Alternatively, and for numerical feasibility, one can consider long-range hopping of electrons on a two-dimensional lattice, i.e., the `SLAC fermion' approach originally proposed in Ref.\cite{drell1976strong}. Specifically, we consider the Hamiltonian originally introduced in Ref.\cite{Lang2019}, where the Hilbert space on site $i$ corresponds to four different species of complex fermions denoted as $c_{i,\tau,\sigma}$ where $\tau=a,b$ is an index that becomes the Dirac-spin at low energies while $\sigma=\uparrow,\downarrow $ denotes flavor index. Correspondingly, we define Pauli matrices $\tau^a$ and $\sigma^a$ with $a = x, y, z$ that act on the Dirac-spin index and the flavor index respectively. The Hamiltonian is given by

\begin{eqnarray}
H & = & H_0 + H_U,\,\,\, \textrm{where} \label{eq:h0} \\
H_0 & = &  \sum_{i,x}\text{i}t(x) c_{i}^\dagger \tau^y c_{i+x}  
 - \sum_{i,y}\i t(y) c_{i}^\dagger \tau^x c_{i+y} , \nonumber \\
H_U & = & \frac{U}{2} \sum_{i,\tau=a,b}(\rho_{i,\tau}-1)^2 \nonumber
\end{eqnarray}
Here $t(r) = \frac{ (-)^r \i\pi t_0 }{ L \sin(\frac{\pi r}{L})}$ with $L$ being the linear system size of the lattice while $\rho_{i,\tau} = \sum_\sigma c_{i,\tau,\sigma}^\dagger c_{i,\tau,\sigma}$ is the fermion density operator for $\tau=a,b$.  From now on, we will set $t_0$ to unity so that all energy scales are measured in units of $t_0$. The global continuous symmetry of $H$ is $SU(2)_{\textrm{flavor}} \times U(1)_{\textrm{charge}}$.

\begin{figure}[t]
	\centering
	\includegraphics[width=0.95\hsize]{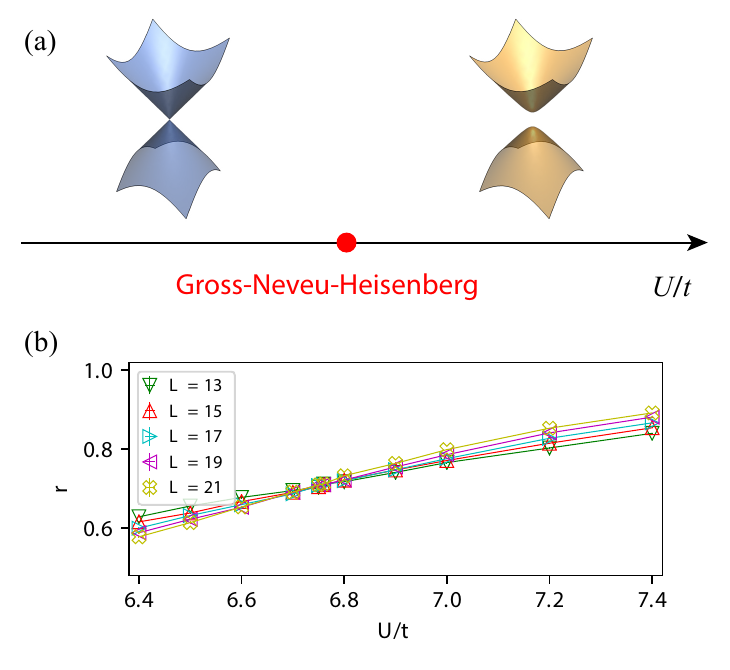}
	\caption{(a) The phase diagram of our model consists of a Dirac semimetal phase separated from a QSH insulator. (b) The critical point between the Dirac semimetal phase and the QSH phase in our lattice model can be located using the crossing point for the correlation ratio $r$ defined as $r=1-\frac{C(\delta{\vec{q}})}{C(\vec{0})}$, where $C(\vec{q})=\frac{1}{L^4}\sum_{i,j}\langle\vec{N}_i\cdot \vec{N}_j\rangle e^{\i \vec{q}\cdot\vec{r}_{ij}}$ is the spin-structure-factor and $\delta\vec{q}=(\frac{2\pi}{L},\frac{2\pi}{L})$. The location of the transition point is consistent with Ref.~\cite{Lang2019}, $U_c/t \approx 6.76$.}
	\label{fig:phasedia}
\end{figure}

The long-range hopping $t(r)$ is precisely the Fourier transform of a dispersion linear in momentum \cite{drell1976strong}, so that in the thermodynamic limit, $H_0$ realizes two flavors of two-component massless Dirac electrons: $H_0 = \sum_{\vec{k},\sigma} c_{\vec{k},\sigma}^\dagger\vec{k} \cdot \vec{\tau} c_{\vec{k},\sigma}$ (see Appendix \ref{sec:slacfermion} for details). One notable aspect is that $H_U$ is \textit{not} Lorentz invariant, and the QMC results in Ref.\cite{Lang2019}  imply that the Lorentz invariance in the Dirac semimetal phase and at the GNH transition is emergent. In addition to the lattice-related symmetries and onsite symmetries corresponding to charge $U(1)$ and flavor $SU(2)$, the model also possesses an onsite anti-unitary symmetry, which we denote as $\mathcal{CT}$, that involves a combination of charge-conjugation and time-reversal: $c_{i,\sigma} \xrightarrow{\mathcal{CT}} \tau^z c_{i,\sigma}^{\dagger}$, $\text{i}  \,(= \sqrt{-1}) \xrightarrow{\mathcal{CT}}-\text{i}$, and time $t \xrightarrow{\mathcal{CT}} -t$. The $\mathcal{CT}$ symmetry is analogous to the one realized at the 2+1-D boundary of a 3+1-D C-I class topological superconductors \cite{wang2014interacting}. Crucially, a combination of $\mathcal{CT}$ and flavor rotation, $c_{i} \rightarrow (\text{i} \sigma^y) \tau^z c_{i}^{\dagger}$, is an antiunitary symmetry that squares to $-\mathds{1}$, and allows one to simulate our model without sign problem ~\cite{Wu04}. Indeed, as mentioned earlier, the phase diagram as a function of $U/t$ has already been mapped out using unbiased QMC simulations in Ref.~\cite{Lang2019}. 

At small $U/t$, the system is in a stable, gapless Dirac semimetal phase. In the continuum limit, the gapless Dirac modes near $\Gamma$ point ($\vec{Q}=(0,0)$) can be written as $c_{\vec{r}} \sim e^{i \vec{Q}\cdot \vec{r}}\Psi$. Although the QMC simulations effectively involve simulating an imaginary time action, for the purposes of discussing symmetries and the duality in the subsequent sections, we will employ a real-time notation (except in Section \ref{sec:rotation} where we study exchange statistics of skyrmions). Defining $\gamma^{0}=\tau^{z}$, $\gamma^{1}=-\text{i}\tau^{y}$, $\gamma^{2}=\text{i}\tau^{x}$, $\overline{\Psi}=\Psi^{\dagger}\gamma^0$, the free part of the Hamiltonian, namely $H_0$, is then described by the standard continuum Dirac Lagrangian $\mathcal{L}_0 = \overline{\Psi}\left( \i \partial_\mu \gamma^{\mu}\right) \Psi$. QMC simulations show that as the interaction strength $U/t$ is increased, the system eventually undergoes a second-order phase transition to a phase with non-zero expectation value $\vec{N} = \langle \overline{\Psi} \vec{\sigma} \Psi \rangle$, see Fig.\ref{fig:phasedia}. In Ref.\cite{Lang2019}, the symmetry-broken phase was referred to as an antiferromagnet. However, we will call it a `quantum spin-Hall insulator' (QSH) since, as discussed below, it exhibits a non-zero spin-Hall response. The phase transition between the semimetal and the QSH phase is expected to be second-order, and can be described by the following field theory: $\mathcal{L} = \overline{\Psi}\left( \i \partial_\mu \gamma^{\mu} + m_N \vec{N}\cdot \vec{\sigma} \right) \Psi + (\partial_\mu \vec{N})^2 + ...$ where the order parameter $\vec{N}$ is normalized as $\vec{N}^2=1$. In addition to sign-problem-free QMC \cite{assaad2013pinning, Toldin14, Otsuka16, Buividovich2018, Lang2019, xu2021competing, otsuka2020dirac}, this critical theory can also be studied using perturbative renormalization group (RG) schemes \cite{rosenstein1993critical,karkkainen1994critical, herbut2006interactions,herbut2009relativistic, Herbut09, janssen2014antiferromagnetic, Zerf2017, Gracey2018, zerf2019critical}.

We will find it useful to couple fermions to probe gauge fields. For most of our discussion, it will suffice to introduce two $U(1)$  gauge fields $A^c$ and $A^s$ that couple to conserved currents $\overline{\Psi}\gamma^\mu \Psi$ and $\overline{\Psi}\gamma^\mu \sigma^z \Psi$  respectively. It is useful to know the transformation properties of these gauge fields, as well as those of various operators relevant to our discussion under discrete symmetry $\mathcal{CT}$, and mirror symmetries $M_x, M_y$ defined as  $M_{x}: c_{\sigma}(x,y)\rightarrow\tau^{x}c_{\sigma}(x,-y), M_{y}: c_{\sigma}(x,y)\rightarrow-\text{i}\tau^{y}c_{\sigma}(-x,y)$, see Table \ref{tab:sym}.  One notices that both charge and flavor currents, i.e. $\overline{\Psi}\gamma^\mu \Psi$ and $\overline{\Psi}\gamma^\mu \sigma^z \Psi$ respectively have the same symmetries as the skyrmion current $j^{\mu}_{\textrm{topo}}= \frac{1}{8\pi}\epsilon^{\mu\nu\lambda} \vec{N} \cdot \partial_\nu \vec{N} \times \partial_\lambda \vec{N}$. One also notices that in addition to `diagonal' Chern-Simons terms such as $A^c dA^c/4\pi$ and $A^s dA^s/4\pi$, even the off-diagonal Chern-Simons term associated with spin-Hall response  $A^c dA^s/4\pi$ is odd under $\mathcal{CT}$. The table also mentions operators involving an internal gauge field $a$ which will be introduced in the next section (see Eq.\eqref{eq:QSH_CP1} below).

\begin{table}[ht]
	\centering
	\begin{tabular}{|c|c|c|c|}
		\hline 
		Operator & $\mathcal{CT}$ & $M_{y}$ & $M_{x}$\\
		\hline 
		\hline 
		$\overline{\Psi}\vec{\sigma}\Psi, \vec{N}$ & - & - & - \\ 
		\hline 
		$\overline{\Psi}\Psi, \vec{N} \cdot \partial^2_x \vec{N} \times \partial^2_y \vec{N}$ & - & - & -\\
		\hline 
		$\overline{\Psi}\gamma^0\Psi, \overline{\Psi}\gamma^0\sigma^z \Psi, A^c_0, A^s_0, \vec{N} \cdot \partial_x \vec{N} \times \partial_y \vec{N}$ & - & + & + \\
		\hline %
		$\overline{\Psi}\gamma^1\Psi, \overline{\Psi}\gamma^1\sigma^z \Psi, A^c_x, A^s_x, \vec{N} \cdot \partial_t \vec{N} \times \partial_y \vec{N}$  & + & - & +\\
		\hline %
		$\overline{\Psi}\gamma^2\Psi, \overline{\Psi}\gamma^2\sigma^z \Psi, A^c_y, A^s_y, \vec{N} \cdot \partial_x \vec{N} \times \partial_t \vec{N}$  & + & + & - \\
		\hline %
		$a_0$ & + & - & - \\
		\hline 
		$a_x$ & - & + & - \\
		\hline 
		$a_y$ & - & - & + \\
		\hline 
		$A^cdA^c, A^s dA^s, A^cdA^s, a da$ & - & - & - \\
		\hline
		$A^cda, A^s da$ & + & + & + \\
		\hline
	\end{tabular}
	\caption{Symmetry transformations of a few operators relevant to our discussion.}
	\label{tab:sym}
\end{table}

\section{Bosonization of Gross-Neveu-Heisenberg transition} \label{sec:gnhduality}
One approach to find a bosonized dual for the GNH transition is to utilize dualities for free fermions for which considerable evidence exists at large-$N$ \cite{maldacena2013constraining, giombi2012chernsimons, aharony2012correlation, aharony2016baryons, seiberg2016dualityweb}, and then append them with appropriate interactions to reach the GNH fixed point. For example, consider the following two Lagrangians with $SU(2)_{\textrm{flavor}} \times U(1)_{\textrm{charge}}$ symmetry:

\bea
\mathcal{L}_{\text{F}} &= & \sum_{a=1}^{2} \overline{\Psi}_a i \slashed{D}_{A}\Psi_a + u (\overline{\Psi} \vec{\sigma} \Psi)^2 - m  \overline{\Psi} \Psi +\textrm{CS}(A)  \label{eq:aharony} \\ 
\mathcal{L}_{\text{B}} & = &  |D_{a+A} \phi|^2  -  \left(\phi^{\dagger}\phi\right)^2 + \textrm{CS}(a) + u \left(\phi^{\dagger} \vec{\sigma} \phi \right)^2  \\
& & - v |\phi^{\dagger} \phi|^3 -r \phi^{\dagger}\phi  \nonumber
\eea
where $CS(X) = \frac{1}{4\pi} \tr \left[XdX - \frac{2i}{3} X^3\right]$ denotes the non-abelian Chern-Simons term for a gauge field $X$, $\Psi_a$ with $a=1, 2$ represents the two flavors of Dirac fermions (Pauli matrices $\vec{\sigma}$ act on the flavor-space) coupled to a background $U(2) = SU(2)_{\textrm{flavor}} \times U(1)_{\textrm{charge}}$ gauge field $A$ in the fundamental representation, and $\phi_a$ with $a = 1,2$ denote $2$ complex scalars that are coupled to a fluctuating $U(N)$ gauge field $a$ as well as the background gauge field $A$ in the fundamental representation. When $u = m = 0$, $\mathcal{L}_F$, the Lagrangian for two flavors of gapless free Dirac fermions, has been conjectured to be dual to $\mathcal{L}_{\text{B}}$, the Wilson-Fisher fixed point Lagrangian of a non-abelian Chern-Simons-matter theory for any value of $N \geq 2$ \cite{maldacena2013constraining, giombi2012chernsimons, aharony2012correlation, aharony2016baryons, seiberg2016dualityweb}. Under this duality, $r \leftrightarrow -m$, i.e., turning on the operator $\pm \phi^{\dagger}\phi $ on the boson side corresponds to turning on the operator $\mp   \overline{\Psi} \Psi$ on the fermion side. For example, giving a positive mass to the boson yields a non-topologically ordered phase (i.e. a unique ground state on a torus) \cite{seiberg2016dualityweb, hsin2016level} and a Hall response $-AdA/4\pi$, matching the fermion theory at negative mass, while giving a negative mass Higgses out the internal gauge field $a$, resulting in a unique, gapped ground state with Hall response $AdA/4\pi$, which again matches with the fermion theory at positive mass. As discussed above, in the presence of time-reversal symmetry (i.e. $m = 0$), tuning the interaction term $u$ in the fermionic Lagrangian $\mathcal{L}_F$ beyond some critical strength drives the GNH transition between the Dirac semimetal and a flavor symmetry broken phase with two Goldstone modes. Due to duality, it is reasonable to expect that the same fixed point can also be reached in the bosonized description $\mathcal{L}_B$ by perturbing the Wilson-Fisher point with a term of the form $ u \left(\phi^{\dagger} \vec{\sigma} \phi \right)^2$ with sufficiently large $u$. Therefore, $\mathcal{L}_B$, at the appropriate fixed point values of the coefficients of  $|\phi|^4$ and $\left(\phi^{\dagger} \vec{\sigma} \phi \right)^2$ can be thought of as the dual description of the GNH transition.

The aforementioned duality proposal for the GNH transition may be worthwhile to analyze in detail, particularly using perturbative methods such as large-$N$ calculations. However, working with non-abelian gauge fields can be a bit challenging. On that note, for the Gross-Neveu-Yukawa phase transition for a single Dirac fermion $\Psi$, where the order parameter corresponds to $\langle \overline{\Psi}\Psi \rangle$, a duality involving only an abelian Chern-Simons-matter theory has been proposed in Ref. \cite{karch2016more}. This duality can be obtained from the `seed duality' between a single Dirac fermion and a single complex scalar coupled to a level-1 Chern-Simons gauge field, with the fermion mass mapping to the boson mass. Using the seed duality one can also argue for a duality between Gross-Neveu-Yukawa phase transition in a two-flavor QED-3 and an SU(2) symmetric $CP^1$ theory, as discussed in Ref.\cite{wang2017deconfined}. This motivates us to ask if there might exist a dual of the GNH transition as well involving only abelian gauge fields. One possible approach is to combine the seed-duality for two different flavors of free Dirac fermions, leading to a dual theory with two complex scalars $z_1, z_2$ coupled to two distinct $U(1)$ gauge fields \cite{potter2017realizing, wang2017deconfined}. One would expect that adding flavor symmetric interactions in such a bosonized description would then drive the GNH transition (in such an approach, the dual of $\overline{\Psi}\sigma^x\Psi$ and $\overline{\Psi}\sigma^y\Psi$ would  involve monopole operators) \cite{footnotewill}. Here we will follow a  different route and consider an alternative candidate duality for the GNH transition, which is  motivated from the quantum numbers of skyrmions in the symmetry-broken phase of our theory.

\underline{Quantum numbers of solitons:} Let us first discuss the effective field theory of the order-parameter-field $\vec{N}$ deep in the ordered phase in the presence of probe gauge fields that couple to the charge and flavor degrees of freedom of the electrons. Coupling SLAC fermions to gauge fields can lead to various inconsistencies \cite{karsten1978axial, karsten1979vacuum, karsten1981lattice} and it is perhaps more appropriate to consider our subsequent discussion in a setup where hopping of fermions is local, e.g., spinful fermions at the boundary of a C-I topological superconductor. After minimal coupling to the probe gauge fields, one may write the effective Lagrangian as

\begin{flalign}
\mathcal{L} = \overline{\Psi}\left( \left( \i \partial_\mu+  A_{\mu}^{c}+ \vec{\sigma}\cdot\vec{A}_{\mu}^{s}\right)\gamma_\mu +  m_N \vec{N}\cdot \vec{\sigma} \right) \Psi + (\partial_\mu \vec{N})^2 \label{eq:GNH}
\end{flalign}
where $A^c$ is a $U(1)$ probe gauge field for the conserved charge, $\vec{A}^s$ is an $SU(2)$ probe gauge field for the conserved flavor, while $m_N$ is a parameter that can be thought of as a Hubbard-Stratonovich parameter for the interaction of the form $(\overline{\Psi} \vec{\sigma} \Psi)^2$. Since we are deep in the ordered phase, we neglect fluctuations of the magnitude of the order-parameter and set $|\vec{N}| = 1$. After integrating out the electrons, one finds the following effective action  \cite{abanov2000theta, Abanov2000}:

\begin{eqnarray}
& & S_{\text{QSH}}[\vec{N}, A^c, A^s]  =  \int d^{2}x\, dt\,\, \left( \frac{|m_N|}{16}\text{tr}(\partial_{\mu}\vec{N})^{2} + \pi H(\vec{N}) \nonumber\right. \\
& & \left. +  j^{\mu}_{\textrm{topo}} A^c_{\mu} + \frac{1}{2\pi} \epsilon^{\mu\nu\lambda} \left( \partial_{\mu} A_{\nu}^{c}\right) \vec{A}_{\lambda}^{s}\cdot\vec{N}\right) \label{eq:abanov}
\end{eqnarray}
Here $H(\vec{N})$ is the Hopf invariant that equals the winding number associated with the homotopy group  $\pi_3(S^2) = \mathbb{Z}$, where the base manifold $S^3$ corresponds to the space-time because one has identified the field configurations of $\vec{N}$ at space-time infinity, while the target manifold $S^2$ corresponds to $\vec{N}$ with $\vec{N}^2 = 1$. The coefficient $\pi$ in front of $H(\vec{N})$ implies that the skyrmions of field $\vec{N}$, whose current in the above equation is denoted as $j^{\mu}_{\textrm{topo}}= \frac{1}{8\pi}\epsilon^{\mu\nu\lambda} \vec{N} \cdot \partial_\nu \vec{N} \times \partial_\lambda \vec{N}$,  have fermionic statistics \cite{Wilczek1983, polyakov1988fermi, dzyaloshinskii1988neutral, abanov2000theta, Abanov2000}. The physical electromagnetic current is given by  $j^{\mu}_c = \left.\frac{\delta S}{\delta A^c_\mu}\right|_{A^c = A^s = 0}$, and Eq.\ref{eq:abanov} implies that  $j^{\mu}_c = j^{\mu}_{\textrm{topo}}$. The time-component of this equation implies that skyrmions carry the same electric charge as the physical electron. This is consistent with the fermionic statistics of the skyrmions and also the fact that the skyrmion density $j^{0}_{\textrm{topo}}$ has the same symmetries as the electron density $\Psi^{\dagger} \Psi$ (Table \ref{tab:sym}). In Sec.\ref{sec:rotation}, we will perform a numerical calculation that provides support for the fermionic exchange statistics of the skyrmions in our model. Finally, the mixed Chern-Simons term between the gauge fields $A^c$ and $\vec{A}^s$ implies that the symmetry-broken phase has a quantized spin-Hall response, and therefore should be identified as a QSH insulator. All of this is quite analogous to the more familiar case of $N_f = 4$ flavors of Dirac fermions (e.g. in graphene), except in that case, one finds bosonic, charge-2 skyrmions whose condensation can lead to a deconfined critical point between a QSH insulator and an s-wave superconductor \cite{grover2008topological, liu2019superconductivity, wang2021doping, hou2022monopole}. We also note that the idea of fermionic skyrmions as induced by a Hopf term was originally discussed in the context of two dimensional antiferromagnets in Refs.\cite{polyakov1988fermi, dzyaloshinskii1988neutral}.  However, as later shown, such a possibility does not occur in a strictly two-dimensional antiferromagnet \cite{fradkin1988topological, wang2014interacting}.

It is useful to re-write the Hopf invariant $H(\vec{N})$ in terms of a Chern-Simons field  \cite{Wilczek1983}. Let us introduce a $CP^1$ representation for the order parameter,  $\vec{N}=z^\dagger \vec{\sigma} z$, where $z$ is a two-component complex vector that satisfies $z^{\dagger}z = 1$. This is a redundant description since $\vec{N}$ is unchanged under the local transformation $z(\vec{r},\tau) \rightarrow e^{i \theta(\vec{r},\tau)} z(\vec{r},\tau) $, which implies that $z$ is coupled to a fluctuating $U(1)$ gauge field $a_\mu$. In this representation, the Hopf invariant, an integer, can be re-written as $H(\vec{N}) = \int_{S^3}\frac{\text{1}}{4\pi^2} a da$ \cite{whitehead1947expression, Wilczek1983}, so that the term $\pi H(\vec{N})$ in the above action precisely has the same form as a Chern-Simons term at level-1.  Further, the skyrmion current is simply given by $j^{\mu}_{\textrm{topo}} = \epsilon^{\mu\nu\lambda}\partial_\nu a_\lambda/2\pi$. In the absence of the $\vec{A}^s$ probe field, the effective field-theory deep in the ordered phase may then be written as: 
\be 
S_{\text{QSH}}  =  \int d^{2}x \,dt\,\, \left( \frac{|(\partial_\mu - \text{i} a_\mu)z|^2}{g^2} + \frac{a dA^c }{2\pi} + \frac{a da}{4\pi} \right) \label{eq:QSH_CP1}
\ee 
where $g^2$ is a coupling constant analogous to $1/|m_N|$ in Eq.\ref{eq:abanov}. The level-1 Chern-Simons term for the gauge field $a$ implies that the `flux-charge composite' operator $z^{\dagger}_\sigma \mathcal{M}$ where $\mathcal{M}$ is a monopole operator that creates $2 \pi$ flux of the gauge field $a$, has the same quantum numbers as the electron creation operator $\Psi^{\dagger}_\sigma$. This composite operator does not carry any gauge charge of the internal gauge field $a$ because both the $CP^1$ bosons $z_\sigma$, as well as  a bare monopole $\mathcal{M}$ carry a unit gauge charge of $a$.  Physically, the action of this composite operator on a given state corresponds to creation of a skyrmionic texture that is bound to an electron. The mixed Chern-Simons term between $a$ and $A^c$ implies that skyrmions carry a unit electric charge.

\begin{figure}[t]
	\centering
	\includegraphics[width=0.9\hsize]{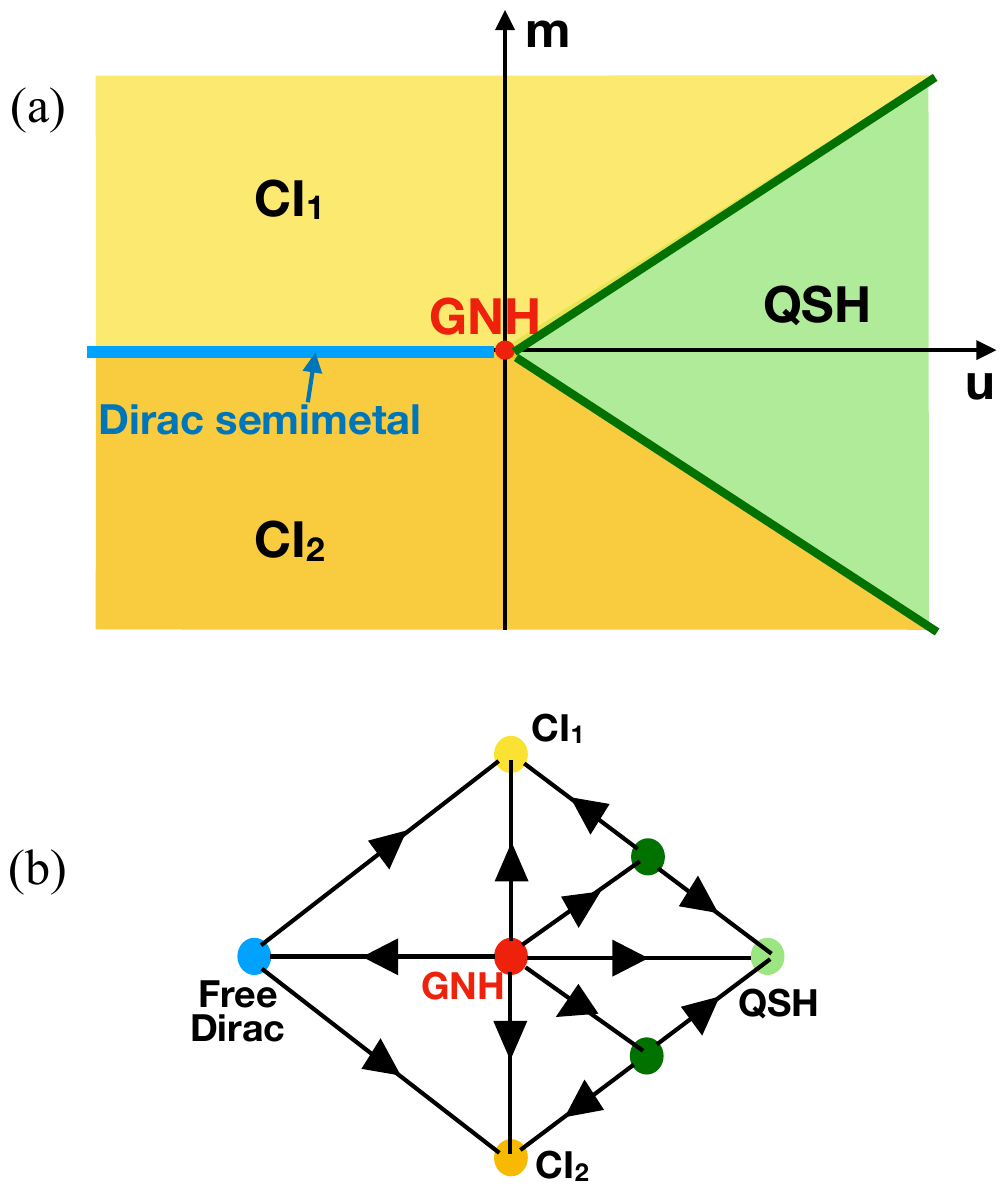}
	\caption{(a) Schematic phase diagram of the Lagrangian in Eq.\eqref{eq:GNH_perturb} in the $(u,m)$ plane, in the vicinity of the GNH critical point. $CI_1$ and $CI_2$ denote the two Chern insulators, QSH denotes the quantum spin Hall phase, GNH denotes the Gross-Neveu-Heisenberg transition between the Dirac semimetal (blue line along the $u$ axis) and the QSH phase. (b) Translation of the phase diagram to an RG flow.}
	\label{fig:phasedia-m-u}
\end{figure}

\underline{Proximate phases:} To motivate a dual field theory for the GNH critical point, it's useful to explore proximate phases that emerge if one perturbs our model Hamiltonian $H$ with terms that break the discrete symmetries $\mathcal{CT}, M_x$ and $M_y$ explicitly. Although we do not know how to simulate the resulting model Hamiltonian due to the fermion sign problem, one may still make a reasonable guess about the phase diagram by considering various limits.  Therefore, consider the following effective Lagrangian in the vicinity of the GNH transition:

\begin{flalign}
\mathcal{L} = \overline{\Psi}\left( \left( \i \partial_\mu+  A_{\mu}^{c}+ \sigma^z A^s_\mu \right)\gamma_\mu - m  \right) \Psi + u (\overline{\Psi} \vec{\sigma} \Psi)^2\label{eq:GNH_perturb}
\end{flalign}
We have supplemented the GNH critical theory with a fermion mass that explicitly breaks the aforementioned discrete symmetries (see Table \ref{tab:sym}) and restricted ourselves to probe gauge field $A^s$ that couple only to the $z$ component of the flavor (we will assume that in the QSH phase, the flavor-symmetry-breaking occurs along the z-direction in the flavor space, so that the spin-rotation symmetry along the z-direction is preserved). Let us write the contribution to the action from the probe fields as $\frac{\sigma^c_{xy}}{4 \pi}  A^c dA^c + \frac{\sigma^{s}_{xy}}{4 \pi}  A^s dA^s + \frac{ \sigma^{sc}_{xy}}{2 \pi}  A^c dA^s$. Integrating out a single flavor of fermion coupled to a U(1) gauge field $b$, a mass $m$ generates a Chern-Simons response $\text{sign}(m) \frac{bdb}{8\pi}$. At $m = 0$, as a function of $u$, the system undergoes the GNH phase transition from the semimetal phase to the QSH phase where $\langle \overline{\Psi}\sigma^z \Psi \rangle \neq 0$. This phase has $\sigma^{c}_{xy} = \sigma^{s}_{xy} = 0$ and $\sigma^{sc}_{xy} = \pm 1$ (the sign of $\sigma^{sc}_{xy}$ depends on the sign of $\langle \overline{\Psi}\sigma^z \Psi\rangle$). On the other hand, when $u = 0$, for $m > 0$, one obtains a flavor-symmetric Chern insulator (which we will denote as `$CI_1$') with $\sigma^c_{xy} = \sigma^s_{xy} = 1$ and $\sigma^{sc}_{xy} = 0$, while for $m < 0$, one obtains a flavor-symmetric Chern insulator  (`$CI_2$') with $\sigma^c_{xy} = \sigma^s_{xy} = -1$ and $\sigma^{sc}_{xy} = 0$. \textit{Assuming} that the phase diagram consists of just these three stable phases, we schematically expect a  phase diagram shown in Fig.\ref{fig:phasedia-m-u}. We propose the following field theory for the transition between the QSH and $CI_2$:

\begin{eqnarray}
S & = & \int d^{2}x \,dt\,\, \left( |(\partial_\mu - \text{i}  a_\mu + \text{i}  A_\mu^s \sigma^z)z|^2 - r |z|^2 +  \frac{a dA^c }{2\pi}  \right.+ \nonumber \\
& & \left. \frac{ a da }{4\pi} +  u(z^{\dagger} \vec{\sigma} z)^2 - \frac{A^s dA^s}{4\pi}  \right)\label{eq:CP1CS}
\end{eqnarray}
where now $z$ is a two-component complex scalar \textit{without} the constraint $z^{\dagger}z = 1$, and $\sqrt{r}$ is the mass for this scalar. Note that $(z^{\dagger} \vec{\sigma} z)^2 = |z^{\dagger}z|^2$. The transition from the $CI_2$ to the QSH phase is driven by changing the sign of $r$. When $ r \ll 0$ or when $u \gg |r|$, we expect that $z$ condenses leading to spontaneous symmetry breaking of the flavor $SU(2)$ down to $U(1)$, resulting in the QSH phase, with $\langle z^{\dagger} \vec{\sigma} z\rangle \neq 0$, and two Goldstone modes. Deep in this phase, if one neglects the fluctuations of $|z|$, one recovers the effective action discussed above using gradient expansion, Eq.\eqref{eq:abanov}, or equivalently Eq.\eqref{eq:QSH_CP1}. Choosing $\langle z_1 \rangle \neq 0$ in this  phase, one finds $a = A^s$ due to the Higgs effect. This correctly reproduces the Hall response  $\sigma^{c}_{xy} = \sigma^{s}_{xy} = 0$ and $\sigma^{sc}_{xy} = 1$ of the QSH phase. On the other hand when $r \gg 0$ and $ r \gg u$, the field $z$ will be gapped and one may integrate it out. After solving for the equations of motion for the gauge fields, one finds $\sigma^c_{xy} = \sigma^s_{xy} = -1$ and $\sigma^{sc}_{xy} = 0$, which we then identify as $CI_2$. One may similarly describe the phase transition between the QSH phase and the $CI_1$ phase by writing down a  similar action where the sign of the $ada$ and $A^s dA^s$ terms are reversed, and one chooses $\langle z_2\rangle \neq 0$.

Above we haven't specified the relation between the fermion mass $m$ in Eq.\eqref{eq:GNH_perturb} and boson mass $\sqrt{r}$ in Eq.\eqref{eq:CP1CS}.  Here we simply mention that at a fixed interaction strength $u$, the $CI_2$ to QSH transition can be accessed by increasing $m$ in the fermionic description  (see Fig.\ref{fig:phasedia-m-u}), and by decreasing $r$ in the bosonized description, which is somewhat similar to standard bosonization dualities. We will elaborate on our understanding and potential issues in more detail below.

\underline{A dual of the GNH transition:} The aforementioned theory for the transition between QSH to Chern insulator (Eq.\eqref{eq:CP1CS}), and the phase diagram (Fig.\ref{fig:phasedia-m-u}) motivates us to conjecture that the GNH theory written in terms of fermions and the order-parameter field, i.e.,

\be 
 \mathcal{L} = \overline{\Psi}\left( \left( \i \partial_\mu+  A_{\mu}^{c}+ \sigma^z {A}_{\mu}^{s}\right)\gamma_\mu +  m_N \vec{N}\cdot \vec{\sigma} \right) \Psi + (\partial_\mu \vec{N})^2 \nonumber
\ee 
is dual to the following theory written in terms of a two-component complex scalar $z$ and a dynamic, compact $U(1)$ gauge field $a$:
\begin{eqnarray}
& & \mathcal{L} = \left( |(\partial_\mu - \text{i}  a_\mu + \text{i}  A_\mu^s \sigma^z)z|^2 +  \frac{a dA^c }{2\pi}  + \frac{ a da }{4\pi} +  u(z^{\dagger} \vec{\sigma} z)^2 \right.\nonumber \\
& & \left. - v |z^{\dagger} z|^3- \frac{A^s dA^s}{4\pi}  \right)\label{eq:CP1CS_tricrit}
\end{eqnarray}

\noindent Since we are interested in the tricritical point, one needs to keep terms upto $|z|^6$ in  Eq.\eqref{eq:CP1CS_tricrit}. This ensures that the symmetry broken phase obtained by changing the sign of $u$ has a well-defined minima for the order parameter $|z|$ \cite{cardy1996scaling}. Higher order terms are not expected to be relevant.
The Lagrangian in Eq.\eqref{eq:CP1CS_tricrit} has identical field content and similar form as the one for the QSH to Chern insulator transition (Eq.\eqref{eq:CP1CS}), except that scalar mass $\sqrt{r} = 0$. We require that the scalar mass $|z|^2$ is not allowed by the symmetry $\mathcal{CT}$. We will discuss justification for imposing this requirement below. This  suggests a single parameter ($= u$) tuned  transition between the QSH phase and a gapless phase without any obvious instabilities that hosts a gauge-neutral (with respect to $a$) field $z^{\dagger}_\sigma \mathcal{M}$ with the quantum numbers of the electron. We posit that the latter phase corresponds to the gapless Dirac semimetal. This suggestion for the dual of Dirac fermion is somewhat similar to that proposed in Ref.\cite{polyakov1988fermi,dzyaloshinskii1988neutral}, although we do not know any controlled calculation, or a known duality that justifies this assumption. Nonetheless, assuming that such an identification is correct, and that there is a unique universality for the phase transition between the Dirac semimetal and the QSH phase, we identify the tricritical theory with the standard GNH transition. Although this is a tricritical point from the perspective of the Chern-Simons-matter theory in Eq.\eqref{eq:CP1CS}, it is a single-parameter-tuned transition when the boson mass $|z|^2$ is prohibited (we assume $ v > 0$). This is reminiscent of other Bose-Fermi dualities where a Gross-Neveu-type theory maps to a tricritical theory of bosons coupled to gauge fields \cite{maldacena2013constraining, giombi2012chernsimons, aharony2012correlation, aharony2016baryons, seiberg2016dualityweb, karch2016more} although our understanding of the theory in Eq.\eqref{eq:CP1CS_tricrit} is comparatively limited.  A heuristic picture for the transition is as follows. The aforementioned flux-charge composite $z^{\dagger}_\sigma \mathcal{M}$ is gapless in the semimetal phase, while it is gapped out in the QSH  phase. In the QSH phase, it carries the same quantum numbers as the electron, as discussed above and has the interpretation of an electron bound to a skyrmionic texture (see the discussion following Eq.\ref{eq:QSH_CP1}). Therefore, closing the gap to the flux-charge composite is tantamount to closing the electron gap. This suggests that the standard GNH Lagrangian (Eq.\eqref{eq:GNH}) is dual to the Chern-Simons-matter theory in Eq.\eqref{eq:CP1CS_tricrit}. In the following, we will explore consequences and potential issues related to this duality conjecture.

Above, we already identified the electron creation operator with the flux-charge composite  $z^{\dagger}_\sigma \mathcal{M}$, and the topological current $ j^{\mu}_{\textrm{topo}}$ with the electromagnetic current $\overline{\Psi} \gamma_\mu \Psi$. One may be inclined to identify the negative of electron mass $-\overline{\Psi} \Psi$ with the boson mass $z^{\dagger} z$, analogous to other Bose-Fermi dualities involving a Chern-Simons term \cite{seiberg2016dualityweb, karch2016particlevortex}. Heuristically, at the level of semiclassical equation of motion for gauge field $a$, $\textrm{Im}(z^{\dagger} \partial_\mu z) + 2 a_\mu z^{\dagger} z + \epsilon_{\mu \nu \lambda} \partial_\nu a_\lambda/2\pi = 0$, which suggests that the operator $z^{\dagger}z$ has the same symmetries as $ada$, which is odd under $\mathcal{CT}$, see Table \ref{tab:sym}. This is also natural from the perspective of the phase diagram in the vicinity of the GNH transition where  $|z|^2$ acts as a tuning parameter for phase transitions (e.g. between QSH and $CI_1$) that are accompanied by a change in the Hall response, as discussed above. However, such an identification does not quite work. When the coefficient $r$ of $z^{\dagger}z$ is large and positive, the $z$ fields have a mass gap, and one finds a Hall response for the probe fields which is consistent with the Chern-insulator phase $CI_2$. This indeed matches with the Hall response in the GNH theory (Eq.\eqref{eq:GNH_perturb}) when the fermion mass $m \ll 0$. However, when the coefficient of $z^{\dagger} z$ is large and negative, one expects to obtain the QSH phase with no charge Hall response and two Goldstone modes. In contrast, in the fermionic theory (Eq.\eqref{eq:GNH_perturb}), reversing the sign of the mass simply reverses the sign of the Hall conductance, and one obtains the $CI_1$ phase.  We don't have a satisfactory resolution to this issue (as an aside, such an issue does not arise if one considers aforementioned duals of GNH that are based on standard Bose-Fermi dualities, such as Eq.\eqref{eq:aharony}, or the one involving two complex scalars coupled to two abelian gauge-fields). A guess for the dual of the fermion mass $\overline{\Psi}\Psi$ is the topological mass term that drives the bosonic integer quantum Hall transition for the $z$ fields between a trivial gapped phase of $z$ bosons, and a non-trivial phase where $z$ bosons are in an integer quantum Hall state with Hall conductance of two. Such a transition will be accompanied by a change in the sign of the $ada/4\pi$ term in Eq.\eqref{eq:CP1CS_tricrit}, resulting in a change in the Hall conductance of our original fermions. Such an identification would be analogous to that obtained for the standard particle-vortex applied to two flavors of Dirac fermions \cite{potter2017realizing,wang2017deconfined}. However, we do not know how to write down such a mass term explicitly in terms of complex scalars $z$.

Similarly, it is not clear to us how to write down the dual of the boson mass $z^{\dagger} z$ under the proposed duality. One naive possibility is that perhaps it corresponds to a linear combination of the two relevant operators at the transition, i.e., $z^{\dagger} z \sim \alpha \overline{\Psi} \Psi   + \beta  (\overline{\Psi} \vec{\sigma} \Psi)^2$, where $\alpha, \beta$ are O(1) numbers. Such an identification would imply that $z^{\dagger}z$ is still prohibited at the GNH transition due to the discrete symmetries, but it is neither even nor odd under these symmetries. As one tunes the coefficient of the $z^{\dagger} z$ term, one moves along a line with slope $\frac{\alpha}{\beta}$ in the phase diagram in the $(u,m)$ plane. An appropriate choice of $\alpha$ and $\beta$ would then be consistent with the requirement that one obtains the QSH phase for $r \ll 0$ and a Chern insulator for $r \gg 0$. On the other hand, when $|r| \ll 1$, so that one is in the scaling regime corresponding to the GNH critical point, $z^{\dagger}z$ will effectively correspond to the operator that has the lower scaling dimension out of $\overline{\Psi} \Psi$ and $(\overline{\Psi} \vec{\sigma} \Psi)^2$ at the GNH critical point (assuming they have different scaling dimensions). Ref.\cite{Lang2019} found scaling dimension of $(\overline{\Psi} \vec{\sigma} \Psi)^2$, $\Delta_{u} \approx 1.98(1)$, and our numerics discussed in the next section found the scaling dimension of $\overline{\Psi}\Psi$ to be $\Delta_m \approx 2.2(3)$. Therefore, error bars preclude a definitive conclusion on which of them is larger. At large $N$, $\Delta_m = 2 + c/N$ where $c > 0$ \cite{zerf2019critical,footnotejoseph} which, in light of the QMC results, is suggestive that $\Delta_u < \Delta_m$. If so, then in the regime $ |r| \ll 1$, for one sign of $r$ the GNH critical point will be unstable towards QSH mass opening (since $(\overline{\Psi} \vec{\sigma} \Psi)^2$ will dominate $\overline{\Psi}\Psi$), while for the opposite sign of $r$, at the leading order, there will be no mass opening while the subleading  term proportional to $\overline{\Psi} \Psi$ will lead to a Chern-insulator-type mass opening.

\section{QMC results and comparison with proposed dual} \label{sec:qmc_results}

In the last section, we discussed two different phase transitions. The first phase transition we discussed is between the QSH phase and the Chern insulator phase. We argued that this transition is described by the field theory in Eq.\eqref{eq:CP1CS}. Although one can estimate the scaling dimensions of various operators for this transition within a  large-$N$ RG calculation \cite{wen1993transitions,park1992critical}, the Hamiltonian/action for this transition  (using either the fermionic description or the bosonic description) suffers from sign problem, and therefore, we are unable to make any numerical comparison with the field theory results. The second phase transition we discussed, which is the main focus of this work, is the GNH transition between the Dirac semimetal and the QSH phase. We argued that it admits a dual description as a tricritical Chern-Simons matter theory (Eq.\eqref{eq:CP1CS_tricrit}). For this transition, although there is a sign-problem in the conjectured bosonic description (Eq.\eqref{eq:CP1CS_tricrit}), there is no sign problem in the fermionic description \cite{Lang2019}. This offers an opportunity to potentially compare universal exponents obtained from the QMC with those obtained from large-$N$ RG calculations. One  potential obstacle with such a comparison is that not much is known about the tricritical theory directly using large-$N$ methods. However, as we will discuss below, in the large-$N$ limit, the critical value of the interaction strength $u_c$ at the bicritical point is very small, which suggests that in the large-$N$ limit, the exponents of the bicritical point are likely close to those for the tricritical point. At the very least, such a  comparison can be a starting point for future investigations of the proposed duality.
We will also compare QMC exponents with the mean-field theory for the tricritical point.

\textbf{Scaling dimension of fermion operator:}
The conjectured duality predicts that the scaling dimension of the electron creation operator in the GNH theory corresponds to the (dressed) monopole operator that creates  $2\pi$ flux in the tricritical Chern-Simons-matter theory, Eq.\eqref{eq:CP1CS_tricrit}. Based on the QMC calculations in Ref.\cite{Lang2019}, the electron creation operator $\Psi^{\dagger}$ has a scaling dimension of approximately 1.09(1), which is also consistent with our QMC simulations, and is also quite close to the large-$N$ result on the GNH theory upto $O(1/N^3)$ \cite{Gracey2018}, which yields an approximate scaling dimension of approximately 1.10. The monopole scaling dimension in the standard bicritical theory (Eq.\eqref{eq:CP1CS}) has been performed in Ref.\cite{chester2018monopole}. It was found that when the ratio $k/N = 1/2$ where $k$ is the level of the Chern-Simons and $N$ is the number of complex scalars, at large $N$, the saddle point value of the critical interaction $u_c$ at the bicritical theory is almost zero ($\approx 0.02$) for a dressed monopole of flux $2 \pi$ \cite{chester2018monopole}. Therefore, one expects that the leading large-$N$ result for the scaling dimension of a $2 \pi$ flux monopole in the \textit{tricritical}  theory is close to that in the bicritical theory.  Assuming this is the case, one finds that the scaling dimension of the monopole operator that creates flux $ 2 \pi$ is approximately $0.53 N$ \cite{chester2018monopole}. Therefore, for the problem of our interest, namely $N = 2$, one finds that the scaling dimension of the operator that creates a flux-charge composite dual to the electron is approximately 1.06 at the leading order, which is rather close to the QMC result in the GNH theory.

\textbf{Scaling dimension of charge-2 operator:} Operators that are Lorentz scalars and carry charge-2 of the global $U(1)_{\text{charge}}$ correspond to $4 \pi$ flux dressed monopoles under the duality, and it is instructive to compute their scaling dimensions using QMC as well \cite{thankwilliam}. We consider two-point correlations of two distinct pairing operators, the onsite pairing operator $P_\text{os}(i) = c_{i,a,\uparrow}^{\dagger}c_{i,b,\downarrow}^{\dagger}-c_{i,b,\uparrow}^{\dagger}c_{i,a,\downarrow}^{\dagger}$, and the nearest neighbor pairing operator $P_\text{nn}(i) = \sum_{\delta}\left((c_{i,a,\uparrow}^{\dagger}c_{i+\delta,b,\downarrow}^{\dagger}+c_{i,a,\downarrow}^{\dagger}c_{i+\delta,b,\uparrow}^{\dagger})-a\leftrightarrow b\right)$, where $\delta$ denotes the four nearest neighbors on a square lattice, that is, $\pm\hat{x},\pm\hat{y}$. One may verify that both of these are Lorentz scalars (i.e. Dirac-spin singlet). The scaling dimensions for either of these operators are close to each other: $\Delta_{P_\text{os}} \approx 2.5(2)$, and  $\Delta_{P_\text{nn}} \approx 2.6(1)$, see Fig.\ref{fig:all}(a) in the main text, and Figs.\ref{fig:ucppntscaling},\ref{fig:ucppotscaling} in Appendix \ref{app:scaling}. Assuming our duality conjecture is correct, this number should be compared with the scaling dimension of the dressed $4 \pi$ monopole in the tricritical theory, Eq.\eqref{eq:CP1CS_tricrit}. One again expects that the leading large-$N$ result is close to the one in the bicritical theory obtained in \cite{chester2018monopole}, since the saddle point value of the critical interaction $u_c$ for this calculation at the bicritical theory is again very small ($\approx 0.05$), see Ref.\cite{chester2018monopole}. The leading order result for the $4 \pi$ monopole in the bicritical theory at $N = 2$ is approximately $\Delta_{4\pi\, \textrm{monopole}} \approx 2.69$ \cite{chester2018monopole}, which is again close to our QMC estimate.

\begin{figure}[t]
	\centering
	\includegraphics[width=1.05\hsize]{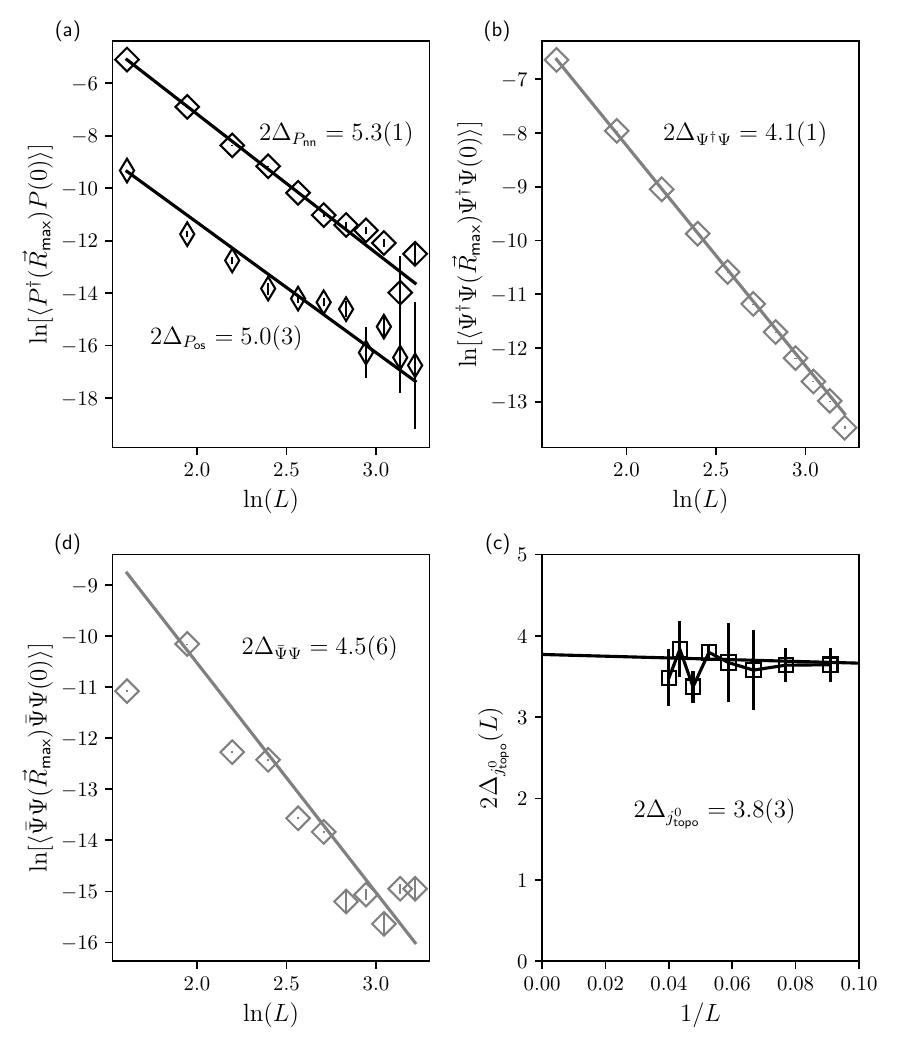}
	\caption{Measurement of the scaling dimension of various operators at the GNH quantum critical point ($U_c/t \approx 6.76$). The data in (a), (b) and (d) are obtained from equal-(imaginary) time unequal-space correlations, while (c) is based on unequal-(imaginary) time, equal-space correlations. The unequal-space correlations are more accurate than the unequal-time since the latter requires two steps of fitting: we first perform a power-law fit for the imaginary time decay at a fixed system size $L$, and then further perform a $1/L$ extrapolation to the thermodynamic limit. The power law fitting of $\Psi^\dagger\Psi$ correlations, $\overline{\Psi}\Psi$ correlations, $P_\text{nn}$ correlations, and $P_\text{os}$ correlations at the largest possible separation ($\vec{R}_\text{max} = (\frac{L-1}{2},\frac{L-1}{2})$) with system size $L$ gives $2\Delta_{\Psi^\dagger\Psi} = 4.1(1)$, $2\Delta_{\overline{\Psi}\Psi}=4.5(6)$, $2\Delta_{P_\text{nn}}=5.3(1)$, and $2\Delta_{P_\text{os}}=5.0(3)$ (the corresponding numbers obtained from unequal time correlations are $2\Delta_{\Psi^\dagger\Psi} = 3.2(1)$, $2\Delta_{\overline{\Psi}\Psi}=4.6(1)$, $2\Delta_{P_\text{nn}}=3.8(1)$, and $2\Delta_{P_\text{os}}=4.6(3)$). From unequal-time skyrmion correlations, we find $2\Delta_{j^0_{\text{topo}}} \approx 3.8(3)$. The data quality for this calculation is further limited by the rather challenging nature of the calculation of skyrmion density correlations. Note that data in grey color are from ``density-channel'' calculation, while that in black color are from ``spin-channel'' calculation. See Appendix~\ref{app:scaling} for more details.
	}
	\label{fig:all}
\end{figure}

\textbf{Scaling dimension of electron charge density and skymion density:} As discussed above, the conservation of total electron number is realized as the conservation of the topological current  $ j^{\mu}_{\textrm{topo}}$ in the Chern-Simons-matter theory. Since conserved charges do not acquire any anomalous dimension, this correspondence predicts that $\Delta_{\Psi^{\dagger} \Psi} = \Delta_{j^0_{\textrm{topo}} } = 2$, where  both $\Delta_{\Psi^{\dagger} \Psi}$ and $\Delta_{j^0_{\textrm{topo}} }$  are obtained using QMC simulations in the model Hamiltonian $H$ by looking at the two-point correlations of the electron density $\Psi^{\dagger} \Psi$ and the skyrmion density $ =  \frac{1}{8\pi}\epsilon^{0 \nu\lambda} \vec{N} \cdot \partial_\nu \vec{N} \times \partial_\lambda \vec{N}$ respectively. Numerically, we find that $\Delta_{\Psi^{\dagger} \Psi} \approx 2.0(1)$ while $\Delta_{j^0_{\textrm{topo}} }\approx 1.9(2)$, see Fig.\ref{fig:all} (b),(c) in the main text and Figs.\ref{fig:u0nnscaling},\ref{fig:u0skyxxscaling},\ref{fig:ucnnscaling},\ref{fig:ucskyxxscaling} in the Appendix \ref{app:scaling}. We note that the calculation for the skyrmion density correlations is rather challenging since this correlation function involves a product of twelve fermion creation or annihilation operators. We used a Mathematica code to generate all possible Wick contractions and after the simplification, each such correlation has 2,064,384 terms, where each term involves a product of six single-particle Green's functions. As an aside, the prefactor $C_J$ for the power-law decay, defined as $\Psi^{\dagger}(x) \Psi(x)\Psi^{\dagger} (0)\Psi(0) \sim C_J/x^4$ is also universal, and will take a different value for the GNH fixed point compared to the free-fermion fixed point. However, we do not have the numerical precision to estimate it reliably.

\textbf{Critical exponent $\nu$ for diverging correlation length}: The tuning parameter for the GNH transition is the interaction term $u( \overline{\Psi}\vec{\sigma} \Psi)^2$. In Ref.\cite{Lang2019}, it was found that various quantities are a scaling function of $u L^{1/\nu}$ with $\nu \approx 0.98(1)$. Therefore, the correlation length $\xi$ diverges as $\xi \sim u^{-\nu}$, and the scaling dimension of the operator $(\overline{\Psi}\vec{\sigma} \Psi)^2$ is $3 - 1/\nu \approx 2$. We do not have a large-$N$ estimate for this scaling dimension in the tricritical theory. However, the value obtained from the mean-field theory of the tricritical theory is surprisingly close. In particular, within the mean-field theory, the inverse propagator at momentum $\vec{k}$ for the complex scalar is $(k^2 + u \langle |z|\rangle^2)$, and since $\langle|z|\rangle \sim \sqrt{u/v}$ within mean-field, this implies that the mean-field correlation length exponent $\nu_{\textrm{MF}} =1$. Alternatively, one notes that the scaling dimension of $z$ is 1/2 within mean-field, and that of $(z^{\dagger} \vec{\sigma} z)^2 \sim ( \overline{\Psi} \vec{\sigma} \Psi)^2$ is 2. The close match between the mean-field tricritical theory and the QMC results might well be a coincidence, but it's still worth noting.

\textbf{Scaling dimension of the order parameter $\vec{N}$:}
Both Ref.\cite{Lang2019} and our simulations find the scaling dimension of the order-parameter $\vec{N}$, $\Delta_{\vec{N}} \approx 0.75$, see Fig.\ref{fig:ucspsmscaling} in Appendix \ref{app:scaling}. Although the scaling dimension of $\vec{N}$ in the bicritical theory has been calculated, see Ref.\cite{moon2012exotic} (one finds $\Delta_{\vec{N}} \approx 1.12$ if one sets $N=2$ within the large-$N$ calculation in Ref.\cite{moon2012exotic}), we are not aware of a similar calculation for the tricritical theory. Within the mean-field of the tricritical point, since the order parameter is bilinear in the scalar $z$, $\Delta_{\vec{N}, \textrm{MF}} = 1$.

\textbf{Scaling dimension of fermion mass  $ \overline{\Psi} \Psi$:} To obtain scaling dimension of $ \overline{\Psi} \Psi$, we perform finite size scaling in our QMC simulations of upto system sizes with linear length $L = 25$ and find an approximate scaling dimension $\Delta_{ \overline{\Psi} \Psi} = 2.2(3)$, see Fig.\ref{fig:all} (d) in the main text, and Fig.\ref{fig:ucnnzzscaling} in Appendix \ref{app:scaling}. As discussed above, we do not know the precise form the operator dual to fermion mass $ \overline{\Psi} \Psi$, although naively one might expect that the operator with which it has the largest overlap is the boson mass operator $z^{\dagger} z$ in the tricritical theory. Although we do not know any reliable estimate of $\Delta_{z^{\dagger} z} $ in the tricritical theory, within a large-$N$ calculation for the bicritical Chern-Simons theory, Refs.\cite{park1992critical, wen1993transitions}, at the leading order one finds $\Delta_{z^{\dagger} z} = 2$. Within mean field, this scaling dimension would be 1.

\textbf{Estimates for universal entanglement $F$:} The universal part of quantum entanglement for a circular bipartition, generally denoted as $F$, has been shown to monotonically decrease between two renormalization group fixed points in 2+1-D Lorentz invariant field theories \cite{casini2012, myers2010, myers2011holographic, casini2011towards, casini2012, jafferis2012exact, Jafferis2011, Klebanov11,casini2015mutual}. For various Bose-Fermi dualities that hold true in the large-$N$ limit \cite{giombi2012chernsimons, aharony2012correlation, aharony2016baryons}, the equality between the $F$ on two sides of duality has already been demonstrated in Ref.\cite{giombi2017testing}. A judicious choice of RG flows connecting theories of interest can constrain phase diagrams \cite{grover2014ftheorem}, and could potentially rule out our conjectured duality. Using the monotonicity property, $F_{\textrm{GNH}} > F_\textrm{Dirac semimetal} = 2 F_D $ where $F_D \approx 0.2190$ is the value of $F$ for a single two-component Dirac fermion. This bound motivates one to find an upper bound for the tricritical Chern-Simons-matter theory (Eq.\eqref{eq:CP1CS}), so as to possibly find contradiction with the conjectured duality. However, unlike the standard $O(N)$ Wilson-Fisher fixed point for which the Gaussian fixed point provides an obvious upper bound, here the presence of the Chern-Simons term makes it difficult to find an analogous bound. Therefore we will simply estimate the two sides using results from large-$N$ expansions. Using results from Ref.\cite{Klebanov11}, for an $SU(2)$ GNH fixed point with $N$ doublets of two-component Dirac spinor, $F_{GNH} = 2 N F_D + 3 \zeta(3)/8\pi^2 + O(1/N)$. Substituting $N = 1$, one finds that to this order  $F_{\textrm{GNH}} \approx 0.48$. To estimate $F$ for the tricritical Chern-Simons-matter theory, we use the inequality $F_{\textrm{tricritical CS}} > F_{\textrm{bicritical CS}} $ where  $F_{\textrm{bicritical CS}}$ can be estimated from the large-$N$ results in Ref.\cite{klebanov2012entanglement}. It was found that for the $CP^{N-1}$ theory with a level $k$ Chern-Simons term, $F_{\textrm{bicritical CS}}  = N F_S + \frac{1}{2}\log\left(\sqrt{k^2 + (\pi N /8)^2} \right) + O(1/N)$ where $F_S \approx 0.1276$ is the $F$ for a free complex scalar. Substituting $N = 2, k = 1$, one finds  $F_{\textrm{tricritical CS}} > F_{\textrm{bicritical CS}}  \approx 0.38$, which is not too far from the aforementioned estimate for $F_{GNH}$.

\section{Numerical demonstration of fermionic skyrmions} \label{sec:rotation}

\begin{figure}[t]
	\centering
	\includegraphics[width=0.95\hsize]{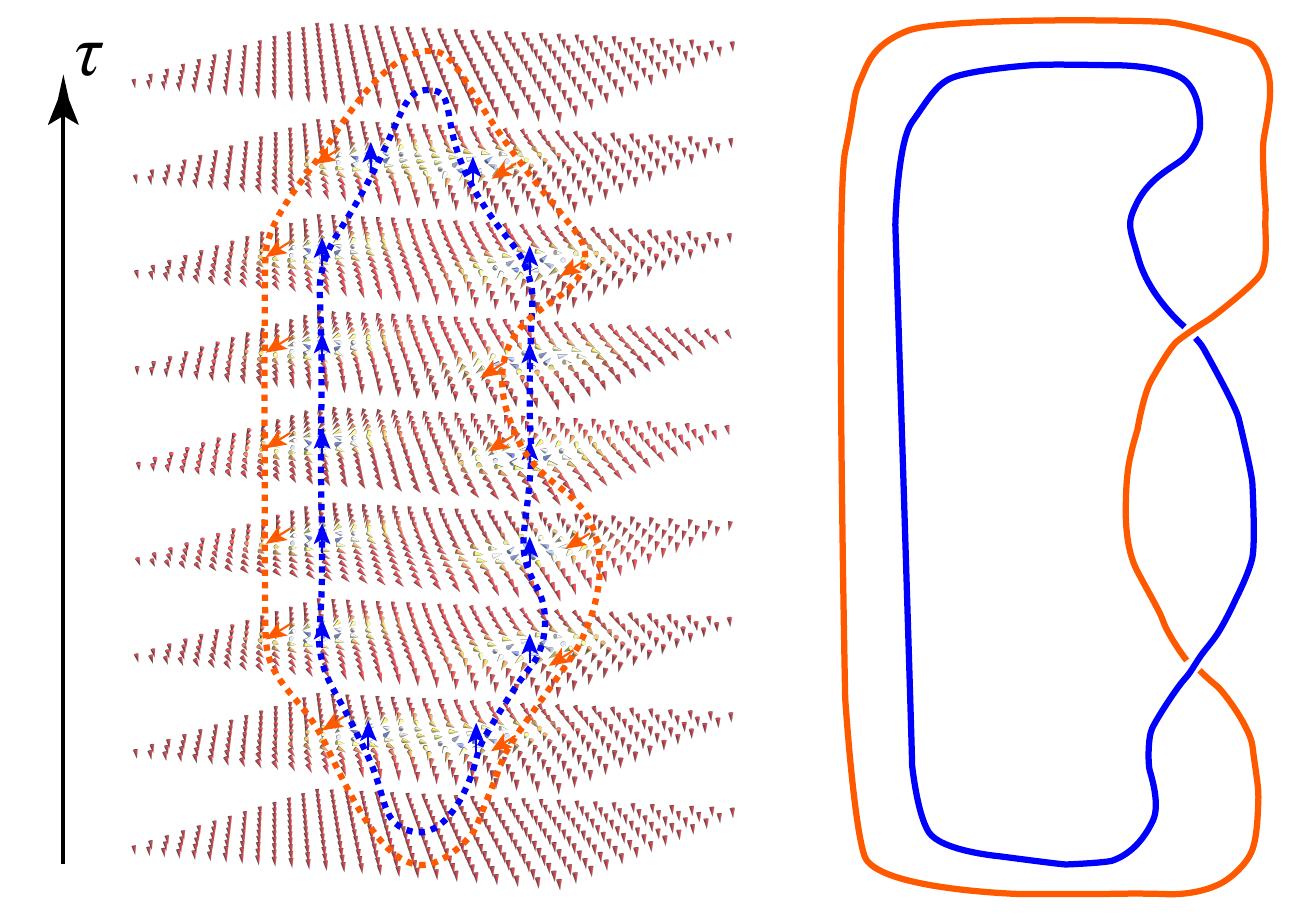}
	\caption{The schematic diagram showing the rotation of skyrmion in a skyrmion-antiskyrmion pair so as to obtain the exchange statistics of a skyrmion.}
	\label{fig:skyrmion}
\end{figure}

To provide evidence for the presence of fermionic skyrmions in the ordered phase, we  consider the imaginary time motion of electrons  in a specific space-time configuration of the order parameter $\vec{N}(\vec{r},\tau)$. In particular, starting with a uniform configuration of $\vec{N}$, we first create a skyrmion-antiskyrmion pair, then separate them, followed by a continuous $2\pi$ rotation of the skyrmion while keeping the antiskyrmion static, and finally bringing them close together and annihilating them, see Fig.\ref{fig:skyrmion} \cite{Wilczek1983}. We also consider a reference path where we rotate the skyrmion from zero to $\pi$ and then back to zero such that the net rotation is zero. We chose  $\vec{N}(\vec{r},\tau)$ so that these two paths lead to the identical contribution to the lattice analog of the `kinetic energy' term  $\int d^{2}x d\tau \frac{|m_N|}{16}\text{tr}(\partial_{\mu}\vec{N})^{2} $ and therefore, differ only in the topological Berry phase picked up during the rotation. Since rotation of a fermion leads to a minus one sign, we expect that the ratio of the imaginary-time partition function for these two paths will be minus one if the skyrmions are indeed fermions. This calculation is implemented in the same SLAC fermion lattice regularization of the GNH model that we used for our QMC simulations discussed above (Eq.~\eqref{eq:h0}). The path-integral corresponding to a configuration $\vec{N}(\vec{r},\tau)$ with $\tau$ ranging from $0$ to $\beta$ is $K(0,\beta)=\int D[\overline{\Psi},\Psi]\mathcal{T}\exp\{- S[\vec{N}]\}$, where $S[\vec{N}]=\int_0^\beta d\tau \int d^2 \vec{r} \overline{\Psi}(\i \slashed{\partial} + \i m_N\vec{N}\cdot \vec{\sigma} ) \Psi $. The skyrmion-antiskyrmion configuration can be generated by setting $\vec{N}=(N_x,N_y,N_z)$, where $N_x=\frac{2\text{Re} W}{1+|W|^2}$, $N_y=\frac{2\text{Im} W}{1+|W|^2}$, and $N_z=\frac{|W|^2-1}{|W|^2+1}$ ~\cite{Nazarov1998}. For the skyrmion-antiskyrmion pair, we set $W(z,\tau)=\frac{a}{z+R(\tau)}-\frac{a}{z-R(\tau)}$, where $z=x+\i y$, $a$ controls the size the skyrmion, and $2R(\tau)$ is the time-dependent seperation between the skyrmion and the antiskyrmion. 
We find a systematic relative sign change for $K(0,\beta)$ associated with the rotated skyrmion configuration and the reference path (unrotated skyrmion) for a wide-range of parameters, including different system sizes, skyrmion size and the maximum separation between the pair (see Appendix~\ref{app:skyrmionrotation} for a detailed discussion), which is consistent with the presence of spin-1/2 skyrmions in our model.

\section{Summary and Discussion} \label{sec:discuss}
In this work we studied a model of interacting fermions that displays Gross-Neveu-Heisenberg transition. Motivated from the quantum numbers of skyrmions, we considered a duality between the standard 2+1-D Gross-Neveu-Heisenberg (GNH) critical point for two flavors of two-component Dirac fermions, and a tricritical Chern-Simons-matter theory with two complex scalars coupled to a level-1 abelian Chern-Simons field (Eq.\eqref{eq:CP1CS_tricrit}). The lattice model we studied was originally introduced and studied in Ref.\cite{Lang2019}, and we obtained new results on the scaling dimensions of various operators in the GNH critical theory, and compared them with the operators in the conjectured dual using available results from various perturbative renormalization group calculations. We also discussed a numerical demonstration of the fermionic statistics of the skyrmions in the ordered phase.

There are several open questions and potential issues pertinent to our proposal. Firstly, we do not know how to show that at weak interactions, and in the absence of scalar mass, the Chern-Simons-matter theory flows to the Dirac semimetal phase. As also emphasized, we do not fully understand the relation between the operators on the two sides of the proposed duality, in contrast to other bosonization dualities. In particular, unlike standard Bose-Fermi dualities, identifying the boson mass with the fermion mass does not quite work, which may indicate that either the duality conjecture is incorrect, or perhaps it is unrelated to known bosonization dualities. It could also be interesting to pursue bosonization of GNH transition using approaches that are better understood, at least within large-$N$, such as Eq.\eqref{eq:aharony}, or the duality for the two complex fermions obtained from the duality between a single complex scalar coupled to a level-1 U(1) Chern-Simons field and a single complex fermion \cite{potter2017realizing,wang2017deconfined,footnotewill}.

Recent large-$N$ calculations indicate that tricritical Chern-Simons theories may have a vacuum instability \cite{di2023vacuum}, and in fact if the estimates from the leading large-$N$ results in Ref.\cite{di2023vacuum} are applied to our case, namely Chern-Simons level $k=1$ and two complex scalars, one would conclude that our theory may not be stable. At the same time, the estimates for the regime of stability obtained from large-$N$ may not be accurate for small values of $N$. For example, large-$N$ calculations on the $CP^{N}$ theories without Chern-Simons term also indicate absence of a second-order transition at small values of $N$ (see, e.g., Refs.\cite{halperin1974first, irkhin19961}), contrary to the numerical evidence of well defined second-order transition at small values of $N$ (see, e.g., Refs.\cite{nahum20113d,nahum2013phase}).

As recently argued \cite{da2022caution}, the long-range hopping associated with SLAC fermion regularization can lead to a gap for the Goldstone modes in the symmetry-broken phase. We did not find any signature of a similar gap at the critical point, which is our focus in this work. As shown in previous works (e.g. Refs.\cite{li2018numerical}), the critical exponents obtained using SLAC regularization are in agreement with those obtained from other approaches, e.g., conformal bootstrap \cite{bobev2015bootstrapping}. Nonetheless, it will be useful to obtain a field-theoretic understanding of the effect of long-range hopping associated with SLAC regularization.

Another direction that may be worth pursuing is to supplement our model with interactions that favor binding of skyrmions and which may therefore result in skyrmionic superconductivity, similar to the scenario discussed in the context of deconfined criticality   in Refs. \cite{grover2008topological, liu2019superconductivity, wang2021doping, hou2022monopole}, or more recently in the context of magic-angle graphene \cite{chatterjee2020symmetry, khalaf2021charged,chatterjee2022skyrmion,liu2022visualizing}.

Finally, if the proposed duality is correct, then it would be fruitful to use it to derive other dualities, e.g., by gauging the probe fields, similar to the derivation of multitude of dualities using a `seed' Bose-Fermi duality \cite{seiberg2016dualityweb, karch2016particlevortex}. For example, if one elevates the probe gauge field $A^c$ in Eq.\eqref{eq:CP1CS_tricrit} to a fluctuating one, then on the fermion side of the duality, one obtains the GNH transition in a two-flavor-QED-3, while on the bosonic side, the gauge field $a$ gets Higgsed and one obtains the tricritical $O(4)$ theory (based on the expectation that the $SU(2) $ symmetry is enlarged to $O(4)$, see, e.g.,\cite{azaria1990nonuniversality,chubukov1994quantum,bonati2021breaking}). The critical exponents for the $O(4)$ tricritical point are essentially mean-field since the interactions are only marginally relevant \cite{riedel1972scaling, riedel1972tricritical,stephen1975logarithmic}. Therefore, this argument is suggestive that the QED-3 GNH transition is dual to simply the O(4) Gaussian fixed-point. We leave further explorations of such implications to the future.

\begin{acknowledgments}
	\emph{Acknowledgments:} 
	The authors thank Shai Chester, A. Liam Fitzpatrick, John Gracey, Thomas Lang, Joseph Maciejko, Nathan Seiberg, Ettore Vicari, William Witczak-Krempa,  Cenke Xu, Yi-Zhuang You, and especially John McGreevy for helpful discussions. We  thank Dachuan Lu for  help with the Mathematica code for Wick contraction of skyrmion-skyrmion correlation. X.Y.X. is sponsored by the National Key R\&D Program of China (Grant No. 2021YFA1401400, No. 2022YFA1402702), the National Natural Science Foundation of China (Grants No. 12274289), Shanghai Pujiang Program under Grant No. 21PJ1407200, Yangyang Development Fund, and startup funds from SJTU. TG is supported by the National Science Foundation under Grant No. DMR-1752417.
\end{acknowledgments}

\newpage
\begin{widetext}
\appendix

\section{SLAC fermion} \label{sec:slacfermion}

\begin{figure}[htp]
\centering
\includegraphics[width=0.6\hsize]{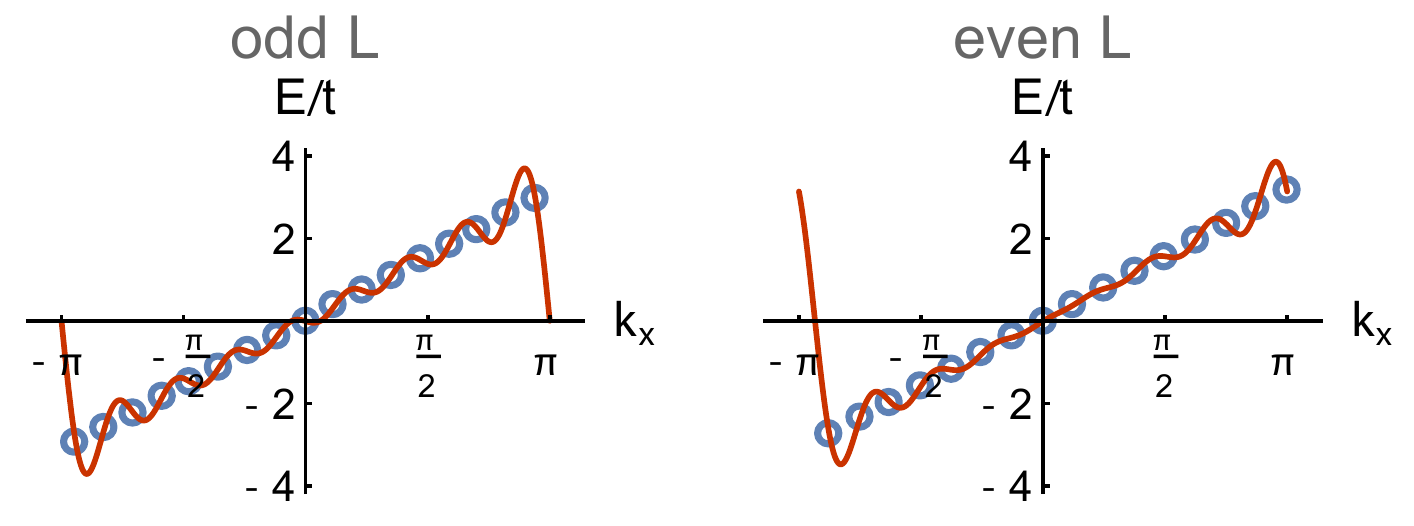}
\caption{SLAC fermion energy-momentum dispersion. The left figure is for odd $L$ ($L=17$) and the right for even $L$ ($L=16$). The blue dots correspond to discrete set of allowed momentum on a lattice, the red lines are the dispersion for continuous $k_x$ on a finite-size lattice. It will approach a straight line in the thermodynamic limit.}
\label{fig:slacdisp}
\end{figure}

As we discussed in the main text, we use SLAC fermion \cite{drell1976strong} to regularize a single four component Dirac cone on a square lattice. The non-interacting part $H_0$ has the form
\begin{equation}
H_0 = \sum_{i,x}t(x) c_{i,a,\sigma}^\dagger c_{i+x,b,\sigma} - \sum_{i,y}\i t(y) c_{i,a,\sigma}^\dagger c_{i+y,b,\sigma} + \hc
\label{eq:h0_append}
\end{equation}
where $L$ is the linear system size of the lattice. The hopping parameter $t(r)$ has the following form for odd $L$
\begin{equation}
t(r)=\begin{cases}
\frac{(-)^{r}\text{i}\pi t}{L\sin(\frac{\pi r}{L})} & r\ne0\\
0 & r=0
\end{cases},
\end{equation}
and the following form for even $L$ 
\begin{align}
t(r) & =\begin{cases}
\frac{(-)^{r}\text{i}\pi te^{-\text{i}\frac{\pi r}{L}}}{L\sin(\frac{\pi r}{L})} & r\ne0\\
\frac{\pi t}{L} & r=0
\end{cases}.
\end{align}
In Fig.~\ref{fig:slacdisp}, we plot the dispersion along $k_x$ direction. The dots corresponding to the discrete set of momenta on the lattice are all located on a straight line.

\section{Details of QMC estimation of scaling dimensions}
\label{app:scaling}
Since our low-energy theory is relativistic, we expect that the dynamical exponent $z=1$ both in the Dirac semimetal phase and at the GNH critical point. In principal, if one has access to arbitrary large system sizes with enough accuracy, one should be able to calculate the scaling dimension of various operators using either the equal-time correlations or unequal-time correlations. However, in practice we find that for some operators, it is easier to estimate their scaling dimension using equal-time, unequal-space correlations, while for others, unequal-time, equal-space yields better estimates.

In Figs.~\ref{fig:u0nnscaling}-\ref{fig:u0skyxxscaling}, we compare the imaginary-time correlation and  the real-space correlation for electron-density operator and the skyrmion-density operator for non-interacting Dirac fermions. The exact value of the scaling dimension for either of these operators is two, and from these figures, we notice that both the imaginary-time correlation as well as the real-space correlation yields an accurate estimate in the thermodynamic limit.

\begin{figure}[htp]
	\centering
	\includegraphics[width=0.9\hsize]{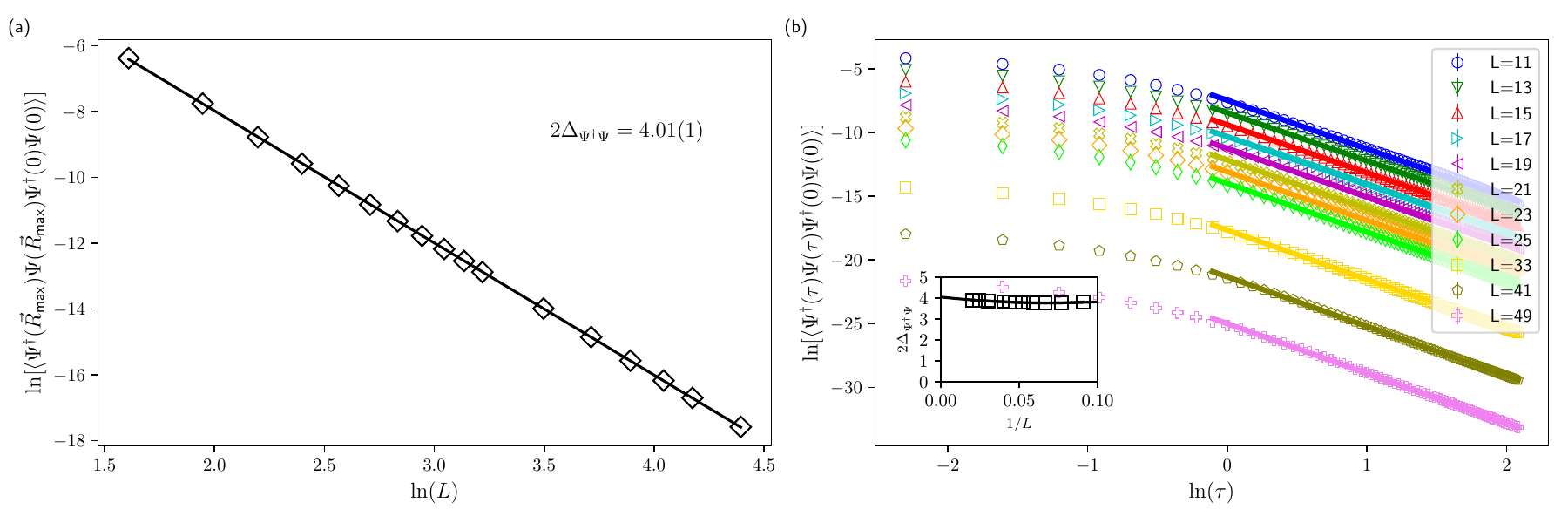}
	\caption{Measurement of scaling dimension of density operator for non-interacting Dirac fermions. (a) Real-space density operator correlation at largest possible separation ($\vec{R}_\text{max} = (\frac{L-1}{2},\frac{L-1}{2})$) for different system size $L$. We obtain $2\Delta_{\Psi^\dagger \Psi} = 4.01(1)$ based on a power-law fitting. (b)Imaginary-time correlation of density operator. We drop the initial 8 points for each $L$ fitting and the fitting range is indicated by a solid line in the figure. Explicitly, the fitting range is $\tau t \in (0.9,8) $.  The inset is a linear extrapolation of $2\Delta_{\Psi^\dagger \Psi}$ with $1/L$, and we get $2\Delta_{\Psi^\dagger \Psi}= 4.05(2)$ in the thermodynamic limit.}
	\label{fig:u0nnscaling}
\end{figure}

\begin{figure}[htp]
	\centering
	\includegraphics[width=0.9\hsize]{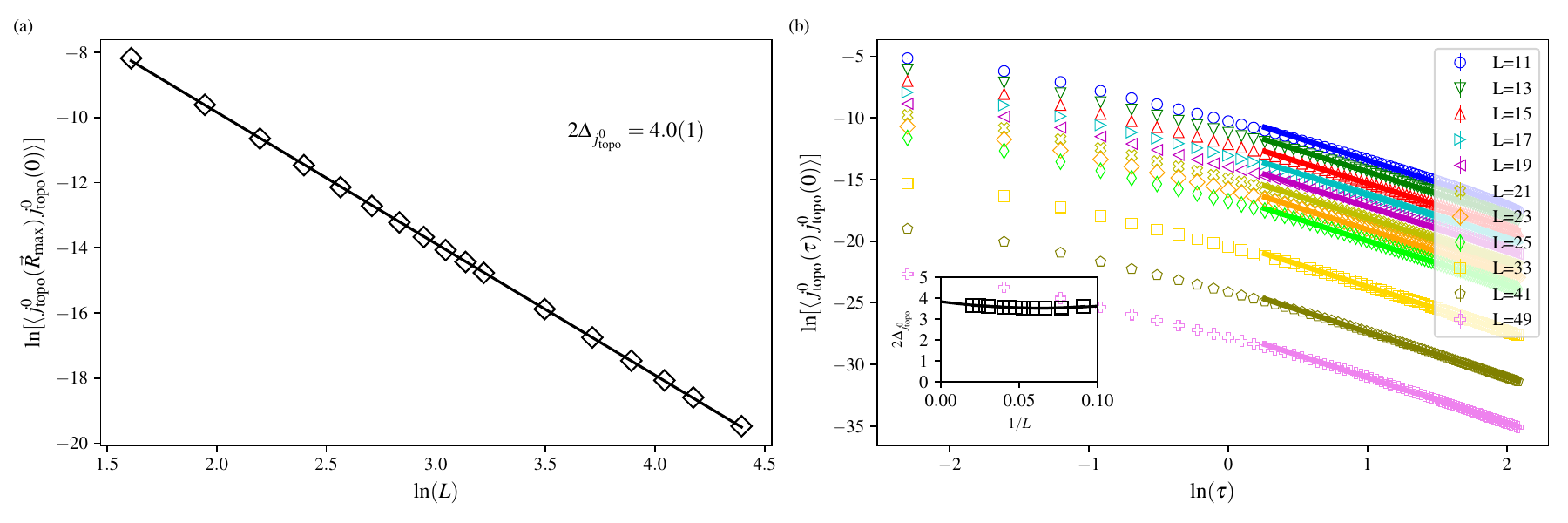}
	\caption{Measurement of scaling dimension of skyrmion-density operator for non-interacting Dirac fermions. (a) Real-space skyrmion-density operator correlation at largest possible separation ($\vec{R}_\text{max} = (\frac{L-1}{2},\frac{L-1}{2})$) for different system size $L$. We obtain $2\Delta_{j^0_{\text{topo}}} = 4.0(1)$ based on a power-law fitting. (b)Imaginary-time correlation of skyrmion-density operator. We drop the initial 8 points for each $L$ fitting and the fitting range is indicated by a solid line in the figure. Explicitly, the fitting range is $\tau t \in (0.9,8) $. The inset is a linear extrapolation of $2\Delta_{j^0_{\text{topo}}}$ with $1/L$, and we get $2\Delta_{j^0_{\text{topo}}}= 3.8(1)$ in the thermodynamic limit.}
	\label{fig:u0skyxxscaling}
\end{figure}

In Figs.~\ref{fig:ucnnscaling}-\ref{fig:ucppotscaling}, we estimate the scaling dimensions of various operators at the GNH critical point. To improve the estimation, we tried two different kinds of Hubbard-Stratonovich transformations (see Appendix.~\ref{app:qmc}), the ``spin-channel" and the ``density-channel". The ``spin-channel" one (denoted by colored points in the figures) has a higher quality of data for unequal-time skyrmion density correlations, and the ``density-channel" one (denoted by grey points in the figures) has a higher quality of data for unequal-time $\bar{\Psi}\Psi$ correlations and $\Psi^\dagger\Psi$ correlations. See Figs.\ref{fig:ucnnscaling},\ref{fig:ucskyxxscaling},\ref{fig:ucnnzzscaling} for details. The calculation of skyrmion-density correlation is particularly challenging, as they involve Wick contractions of a product of twelve fermion operators. With the help of a Mathematica code, we perform the Wick contractions and after the simplification, each two-point correlation of the skyrmion density has 2,064,384 terms, where each term involves a product of six single-particle Green's functions. It appears that the unequal-time, equal-space correlation has a much higher quality than the equal-time, unequal-space correlation, see Fig.~\ref{fig:ucskyxxscaling} for details.

\begin{figure}[htp]
	\centering
	\includegraphics[width=0.9\hsize]{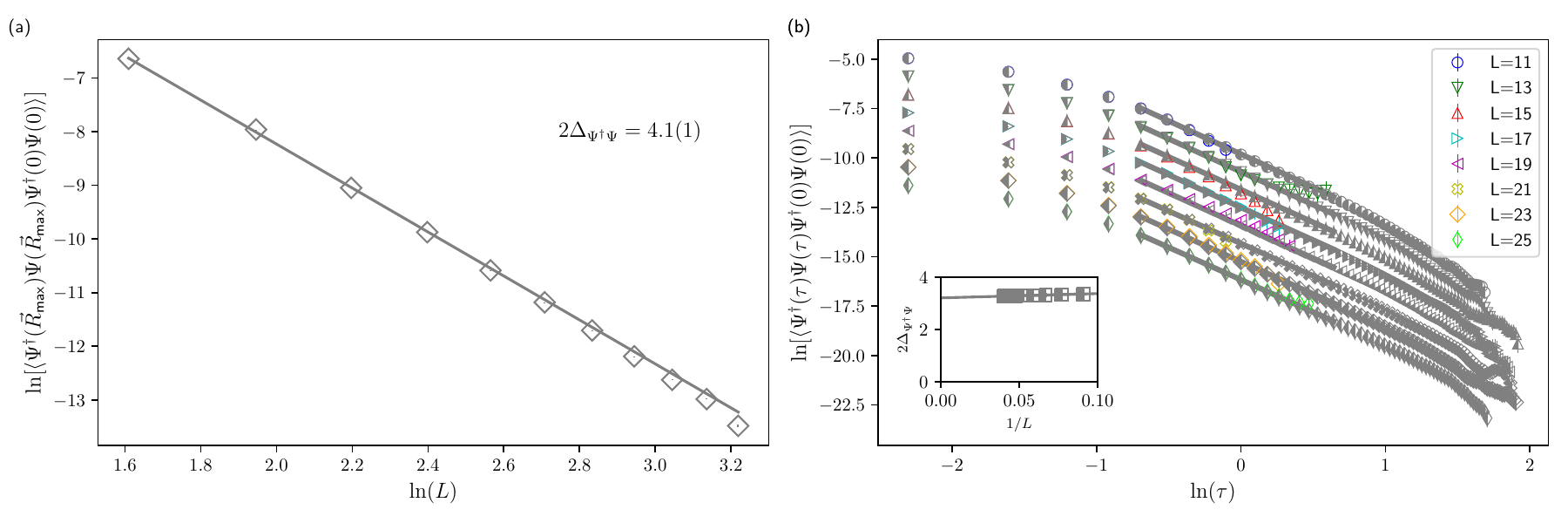}
	\caption{Measurement of the scaling dimension of the density operator at the GNH critical point. (a) Real-space density operator correlation at largest possible separation ($\vec{R}_\text{max} = (\frac{L-1}{2},\frac{L-1}{2})$) for different system size $L$. We obtain $2\Delta_{\Psi^\dagger \Psi} = 4.1(1)$ based on a power-law fitting. (b)Imaginary-time correlation of density operator. We drop the initial 4 points for each $L$ fitting , and the fitting range is indicated by a solid line in the figure. The inset is a linear extrapolation of $2\Delta_{\Psi^\dagger \Psi}$ with $1/L$, we get $2\Delta_{\Psi^\dagger \Psi} = 3.2(1)$ in the thermodynamic limit. Note that the ``spin-channel'' Hubbard-Stratonovich transformation data is denoted by colored points, while the ``density-channel'' one is denoted by grey points. Similar notation is used in the following figures. }
	\label{fig:ucnnscaling}
\end{figure}

\begin{figure}[htp]
	\centering
	\includegraphics[width=0.9\hsize]{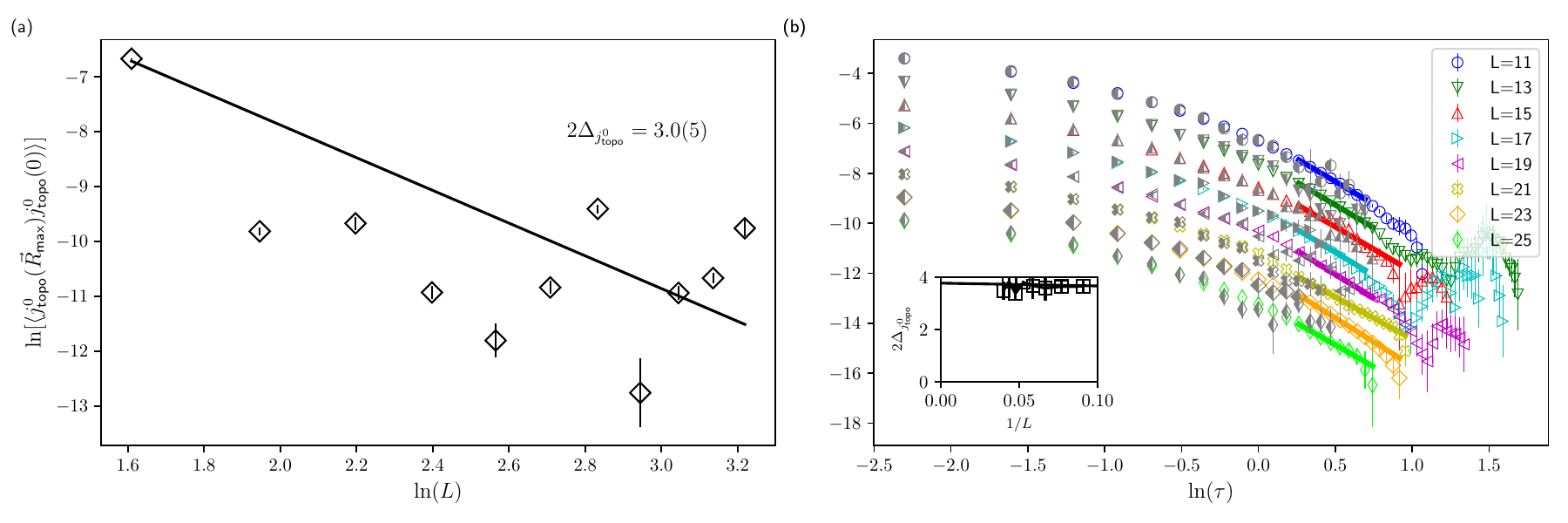}
	\caption{Measurement of the scaling dimension of the skyrmion-density operator at GNH critical point. (a) Real-space skyrmion-density operator correlation at largest possible separation ($\vec{R}_\text{max} = (\frac{L-1}{2},\frac{L-1}{2})$) for different system size $L$. We obtain $2\Delta_{j^0_{\text{topo}}} = 3.0(5)$ based on a power-law fitting. (b)Imaginary time correlation of skyrmion-density operator. We drop several small $\tau$ and large $\tau$ points for each $L$ fitting, and the fitting range is indicated by a solid line in the figure. The inset is a linear extrapolation of $2\Delta_{j^0_{\text{topo}}}$ with $1/L$, and we get $2\Delta_{j^0_{\text{topo}}}= 3.8(3)$ in the thermodynamic limit.}
	\label{fig:ucskyxxscaling}
\end{figure}

\begin{figure}[htp]
	\centering
	\includegraphics[width=0.9\hsize]{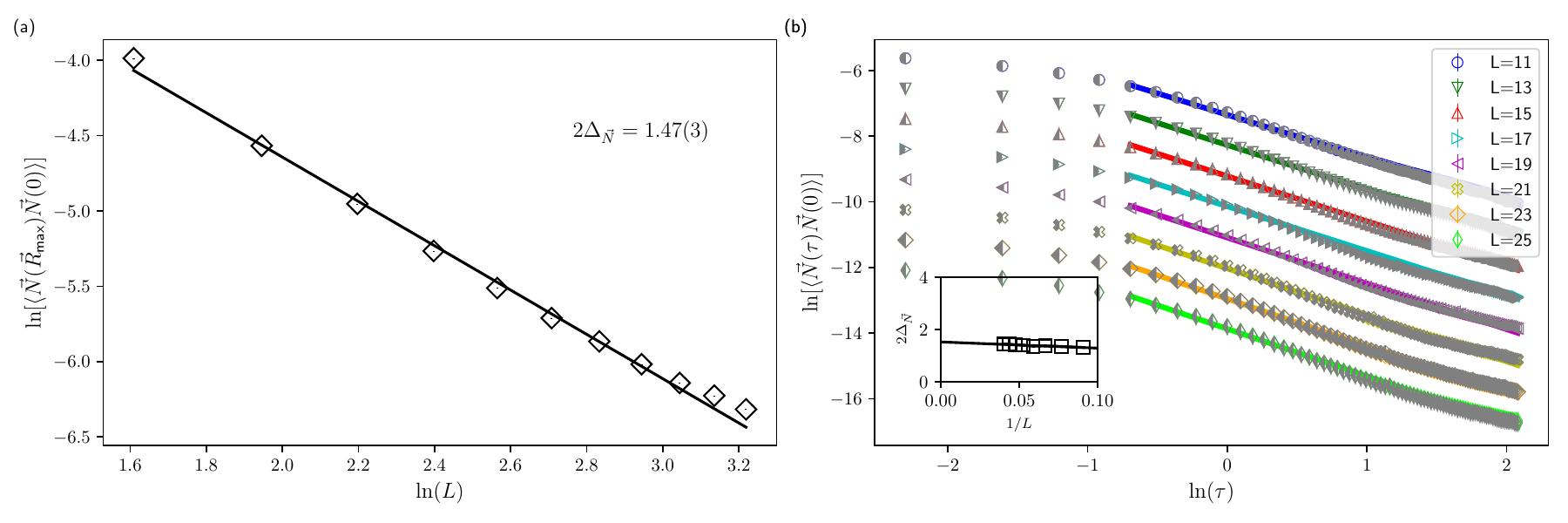}
	\caption{Measurement of the scaling dimension of the QSH order parameter $\vec{N}$ at the GNH critical point. (a) (a) Real-space QSH operator correlation at largest possible separation ($\vec{R}_\text{max} = (\frac{L-1}{2},\frac{L-1}{2})$) for different system size $L$. We obtain $2\Delta_{\vec{N}} = 1.47(3)$ based on a power-law fitting. (b)Imaginary-time correlation of the QSH order parameter. We drop the initial 4 points for each $L$ fitting  and the fitting range is indicated by a solid line in the figure. The inset is a linear extrapolation of $2\Delta_{\vec{N}}$ with $1/L$, and we get for both the ``spin-channel" calculation and ``density-channel" calculation $2\Delta_{\vec{N}} = 1.52(2)$ in the thermodynamic limit, which matches with the previously reported value in Ref.\cite{Lang2019}.}
	\label{fig:ucspsmscaling}
\end{figure}

\begin{figure}[htp]
	\centering
	\includegraphics[width=0.9\hsize]{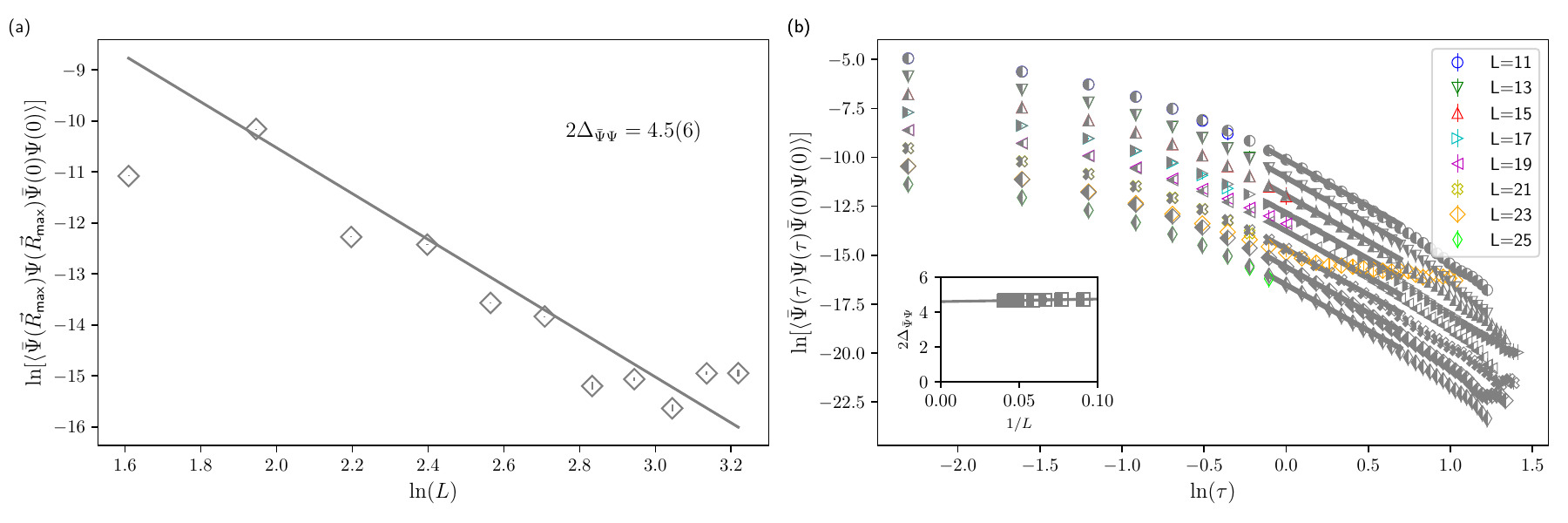}
	\caption{Measurement of the scaling dimension of $\bar{\Psi}\Psi$ operator at the GNH critical point. (a) Real-space $\bar{\Psi}\Psi$ operator correlation at largest possible separation ($\vec{R}_\text{max} = (\frac{L-1}{2},\frac{L-1}{2})$) for different system size $L$. We obtain $2\Delta_{\bar{\Psi}\Psi} = 4.5(6)$ based on a power-law fitting. (b)Imaginary-time correlation of  the  $\bar{\Psi}\Psi$ operator. We drop the initial 8 points for each $L$ fitting and the fitting range is indicated by a solid line in the figure. The inset is a linear extrapolation of the $2\Delta_{\bar{\Psi}\Psi}$ with $1/L$, and we get for ``density-channel" calculation $2\Delta_{\bar{\Psi}\Psi} = 4.6(1)$ in the thermodynamic limit.}
	\label{fig:ucnnzzscaling}
\end{figure}

\begin{figure}[htp]
	\centering
	\includegraphics[width=0.9\hsize]{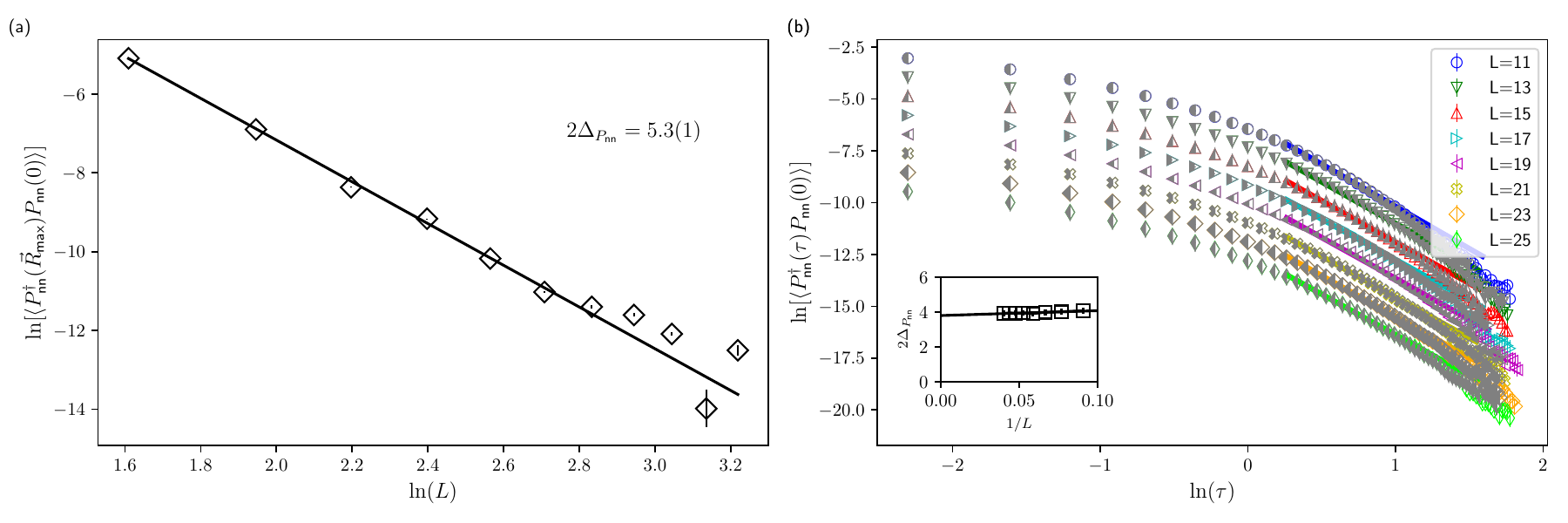}
	\caption{Measurement of the scaling dimension of $P_\text{nn}$ operator at the GNH critical point. (a) Real-space $P_\text{nn}$ operator correlation at largest possible separation ($\vec{R}_\text{max} = (\frac{L-1}{2},\frac{L-1}{2})$) for different system size $L$. We obtain $2\Delta_{P_\text{nn}} = 5.3(1)$ based on a power-law fitting. (b)Imaginary-time correlation of  the  $P_\text{nn}$ operator. We drop the initial 12 points for each $L$ fitting and the fitting range is indicated by a solid line in the figure. The inset is a linear extrapolation of the $2\Delta_{P_\text{nn}}$ with $1/L$, and we get for ``spin-channel" calculation $2\Delta_{P_\text{nn}} = 3.8(1)$ in the thermodynamic limit.}
	\label{fig:ucppntscaling}
\end{figure}

\begin{figure}[htp]
	\centering
	\includegraphics[width=0.9\hsize]{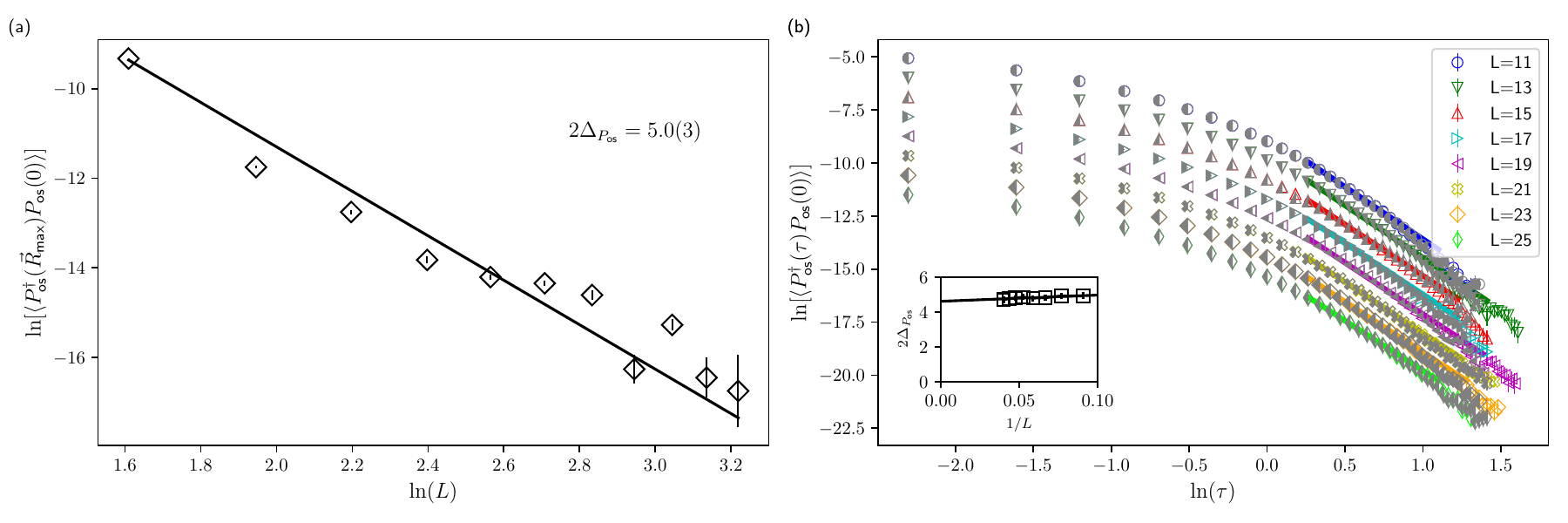}
	\caption{Measurement of the scaling dimension of $P_\text{os}$ operator at the GNH critical point. (a) Real-space $P_\text{os}$ operator correlation at largest possible separation ($\vec{R}_\text{max} = (\frac{L-1}{2},\frac{L-1}{2})$) for different system size $L$. We obtain $2\Delta_{P_\text{os}} = 5.0(3)$ based on a power-law fitting. (b)Imaginary-time correlation of  the  $P_\text{os}$ operator. We drop the initial 12 points for each $L$ fitting and the fitting range is indicated by a solid line in the figure. The inset is a linear extrapolation of $2\Delta_{P_\text{os}}$ with $1/L$, and we get for ``spin-channel" calculation $2\Delta_{P_\text{os}} = 4.6(1)$ in the thermodynamic limit.}
	\label{fig:ucppotscaling}
\end{figure}

\section{Skyrmion rotation calculation}
\label{app:skyrmionrotation}
As discussed in the main text (see Sec.\ref{sec:rotation}), conceptually we consider adiabatic motion of electrons in the background of a specific space-time configuration of the order parameter $\vec{N}(\vec{r},\tau)$ that corresponds to skyrmion rotation (Fig.\ref{fig:skyrmion}), and compare the phase picked up by the electron with a reference configuration where skyrmion is not rotated. In the actual calculation, we use SLAC fermion to regularize the Dirac fermion on a lattice, and make a Trotter decomposition of the imaginary
time $\beta\equiv L_{\tau}\Delta_{\tau}$, where $\Delta_{\tau}$ is
taken to be very small so as to implement the adiabatic motion. In the calculation, we set $L_\tau=400$, and $\Delta_{\tau}t=0.1$. The
space-time Hamiltonian is written as
\begin{equation}
H[\vec{N}]=H_{0}-m\sum_{i}\vec{N}(\vec{r}_{i},\tau)\cdot c_{i}^{\dagger}\tau^{z}\vec{\sigma}c_{i}\equiv c^{\dagger}h(\tau)c
\end{equation}
where $h(\tau)$ is the coefficient matrix of the space time Hamiltonian
at imaginary time. After tracing out fermions, one obtains
\begin{equation}
K(0,\beta)=\det\left[1+\prod_{l=1}^{L_{\tau}}e^{-\Delta_{\tau}h(l\Delta_{\tau})}\right]
\end{equation}
The skyrmion configuration can be generated by setting $\vec{N}=(N_x,N_y,N_z)$, where $N_x=\frac{2\Re W}{1+|W|^2}$, $N_y=\frac{2\Im W}{1+|W|^2}$, and $N_z=\frac{|W|^2-1}{|W|^2+1}$ ~\cite{Nazarov1998}. For the skyrmion-antiskyrmion pair, we can set $W(z)=\frac{a}{z+R}-\frac{a}{z-R}$, where $z=x+\i y$, $a$ is the size the skyrmion, and $2R$ is the separation of the skyrmion and the antiskyrmion. To describe the separation process we make $R$ to be time-dependent, and similarly, $W(z)$ depends on the time as well to implement the rotation:
\begin{equation}
W(z) = \frac{a}{z+R(\tau)}-\frac{a}{z-R(\tau)}e^{\i \alpha(\tau)}
\end{equation}
During the period when the skyrmion-antiskyrmion is created out of the vacuum and slowly separated, we set $\alpha(\tau)=0$ as $R(\tau)$ changes from zero to $R_0$ in this process. The reverse process of annihilating is also carried out similarly. During the period when skyrmion is being rotated, we set $R(\tau)=R_0$ fixed, and slowly increase $\alpha(\tau)$ from zero to $2\pi$. The rotation process is made very slow by dividing the angle $2\pi$ into 300 small steps. As mentioned above and in Sec.\ref{sec:rotation}, we also consider a reference configuration, where we rotate the skyrmion from zero to $\pi$ and then from $\pi$ back to zero, such that in total there is no rotation. We tried a range of parameters: We consider different sets of system sizes, $\{L_x=4R_0+1, L_y=2R_0+1 \}$ with $R_0=5,6,7,8$. We also considered different values of $a=2,3,4$ corresponding to different sizes for the skyrmion. Larger $a$ is not suitable due to limited total system size we can simulate. Finally, we also considered a different set of mass ratios in the range $0 \lesssim m/t \lesssim 4$. We obtained a relative sign change for the propagator $K(0,\beta)$ corresponding to the rotated skyrmion compared to that of the unrotated one for all sets of $L$ and  $a$ when $1.0 \lesssim m/t \lesssim 2.5$, as shown in Table~\ref{tab:tab2}. For larger $m/t $, we do not find a sign change which may be related to the fact that when $m/t$ becomes large, ultraviolet physics may affect the result of the calculation since the phase stiffness of the $\vec{N}$ is proportional to $|m|$. This provides a numerical demonstration of spin-1/2 skyrmions, at least for a range of parameters.

\begin{table}[h!]
  \centering
  \begin{tabular}{|c|c|c|c|c|c|c|c|c|c|c|c|c|}
    \hline
    \multicolumn{1}{|c|}{} & \multicolumn{3}{c|}{$R_0=5$} & \multicolumn{3}{c|}{$R_0=6$} & \multicolumn{3}{c|}{$R_0=7$} &
\multicolumn{3}{c|}{$R_0=8$}  \\
    \hline
    $m/t$ & $a=2$ & $a=3$ & $a=4$ & $a=2$ & $a=3$ & $a=4$ & $a=2$ & $a=3$ & $a=4$ & $a=2$ & $a=3$ & $a=4$ \\
    \hline
    0.5 & $+$ & $+$ & $+$ & $+$ & $+$ & $-$ & $+$ & $+$ & $-$ & $+$ & $+$ & $-$ \\
    \hline
    0.6 & $+$ & $+$ & $-$ & $+$ & $-$ & $-$ & $+$ & $-$ & $-$ & $+$ & $-$ & $-$ \\
    \hline
    0.7 & $+$ & $-$ & $-$ & $+$ & $-$ & $-$ & $+$ & $-$ & $-$ & $+$ & $-$ & $-$ \\
    \hline
    0.8 & $+$ & $-$ & $-$ & $+$ & $-$ & $-$ & $+$ & $-$ & $-$ & $+$ & $-$ & $-$ \\
    \hline
    1.0 & $-$ & $-$ & $-$ & $-$ & $-$ & $-$ & $-$ & $-$ & $-$ & $-$ & $-$ & $-$ \\
    \hline
    1.5 & $-$ & $-$ & $-$ & $-$ & $-$ & $-$ & $-$ & $-$ & $-$ & $-$ & $-$ & $-$ \\
    \hline
    2.0 & $-$ & $-$ & $-$ & $-$ & $-$ & $-$ & $-$ & $-$ & $-$ & $-$ & $-$ & $-$ \\
    \hline
    2.5 & $-$ & $-$ & $-$ & $-$ & $-$ & $-$ & $-$ & $-$ & $-$ & $-$ & $-$ & $-$ \\
    \hline
    3.0 & $+$ & $-$ & $-$ & $+$ & $+$ & $-$ & $+$ & $+$ & $-$ & $+$ & $+$ & $+$ \\
    \hline
    4.0 & $+$ & $+$ & $+$ & $+$ & $+$ & $+$ & $+$ & $+$ & $+$ & $+$ & $+$ & $+$ \\
    \hline
    5.0 & $+$ & $+$ & $+$ & $+$ & $+$ & $+$ & $+$ & $+$ & $+$ & $+$ & $+$ & $+$ \\
    \hline
  \end{tabular}
  \caption{Relative sign change for $K(0,\beta)$ for different parameters.}
  \label{tab:tab2}
\end{table}

\section{Details of Quantum Monte Carlo calculation}
\label{app:qmc}
We perform projection Quantum Monte Carlo calculation. The observables are written as
\begin{equation}
\langle O \rangle = \frac{\langle \Psi_0| O |\Psi_0 \rangle}{\langle \Psi_0 | \Psi_0 \rangle}
\end{equation}
where $|\Psi_0\rangle$ is the ground state wavefunction, and is obtained via projection
\begin{equation}
|\Psi_0\rangle = e^{-\Theta H} |\Psi_T\rangle
\end{equation}
where $\Theta$ is the projection time, $|\Psi_T\rangle$ is the trial wavefunction which is set to be the ground state wavefunction of the non-interacting part of $H$. In the calculation, we set $2\Theta t = 60$, which is large enough both for the equal-time and dynamical calculations. The trotter decomposition step is set as $\Delta_\tau t = 0.1$. To deal with the interaction, we perform a symmetric trotter decomposition,
\begin{equation}
e^{-\Delta_\tau(H_0+H_U)} \approx e^{-\frac{1}{2}\Delta_\tau H_0}e^{-\Delta_\tau H_U}e^{-\frac{1}{2}\Delta_\tau H_0}
\end{equation}
and then considering the following two kinds of Hubbard Stratonovich (HS) transformation. For convinience we rewrite $\tilde{c}_{i,a/b,\uparrow}=c_{i,a/b,\uparrow}$,
$\tilde{c}_{i,a,\downarrow}=c_{i,a,\downarrow}^{\dagger}$, $\tilde{c}_{i,b,\downarrow}=-c_{i,b,\downarrow}^{\dagger}$. The first type of HS transformation is in the so called ``spin-channel".
\begin{equation}
e^{-\frac{U}{2}\Delta_{\tau}(\tilde{\rho}_{i,\tau,\uparrow}-\tilde{\rho}_{i,\tau,\downarrow})^{2}}\approx \frac{1}{4}\sum_{s_{i,\tau}=\pm1,\pm2}\gamma(s_{i,\tau})e^{i\alpha_{1}\eta(s_{i,\tau})(\tilde{\rho}_{i,\tau,\uparrow}-\tilde{\rho}_{i,\tau,\downarrow})}
\end{equation}
where $\alpha_{1}=\sqrt{\frac{U}{2}\Delta_{\tau}}$, $\gamma(\pm1)=1+\sqrt{6}/3$,
$\gamma(\pm2)=1-\sqrt{6}/3$, $\eta(\pm1)=\pm\sqrt{2(3-\sqrt{6})}$,
$\eta(\pm2)=\pm\sqrt{2(3+\sqrt{6})}$.
The second type of HS transformation is in the so called ``density-channel".
\begin{equation}
e^{-\frac{U}{2}\Delta_{\tau}(\tilde{\rho}_{i,\tau,\uparrow}-\tilde{\rho}_{i,\tau,\downarrow})^{2}+\frac{U}{2}\Delta_{\tau}}=\frac{1}{2}\sum_{s_{i,\tau}=\pm1}e^{\alpha_{2}s_{i,\tau}(\tilde{\rho}_{i,\tau,\uparrow}+\tilde{\rho}_{i,\tau,\downarrow}-1)}
\end{equation}
where $\alpha_{2}=\text{acosh}e^{\frac{\Delta_{\tau}U}{2}}$. It turns out the ``spin-channel" calculation is more stable for spin type unequaltime correlations such as skyrmion density correlation, and the ``density-channel" calculation is more stable for density type unequaltime correlations such as $\bar{\Psi}\Psi$ correlations and $\Psi^\dagger \Psi$ correlations.

\end{widetext}
\bibliography{corr}
\bibliographystyle{apsrev4-2}

\end{document}